\documentclass[journal]{IEEEtran}

\ifCLASSINFOpdf
  \usepackage[pdftex]{graphicx}
\else
\fi

\usepackage{booktabs}   % 표 라인 예쁘게
\usepackage{siunitx}    % 단위 표현 (선택)

\usepackage{cite}
\usepackage{array}
\usepackage{arydshln}
\usepackage{comment}
\usepackage{physics}
\usepackage{amsmath}
\usepackage{amssymb}
\usepackage{bm}
\usepackage{xcolor}
\usepackage{tcolorbox}
\usepackage{kotex}
\usepackage{subcaption}  % 최신 subfigure 지원
\usepackage{caption}  % 최신 subfigure 지원

\definecolor{lightolive}{rgb}{0.5, 0.7, 0.2}
\definecolor{darkgreen}{HTML}{006400} % CSS DarkGreen

\usepackage {xcolor}
\usepackage{amsmath,amsthm,amsfonts,amscd}  

\usepackage{algorithmic}
\usepackage{algorithm}

\usepackage{array}
\usepackage{multirow}

\ifCLASSOPTIONcompsoc
  \usepackage[caption=false,font=normalsize,labelfont=sf,textfont=sf]{subfig}
\else
  \usepackage[caption=false,font=footnotesize]{subfig}
\fi

\usepackage{url}

\begin{document}
\title{ 
Coupled Aerodynamic--Electromagnetic Modeling for RCS Estimation of Million-Scale Chaff Clouds with Arbitrarily Curved 3D Geometries
}
%\title{Electromagnetic Wave Scattering of Large-scale Chaff Cloud using Surface Integral Equation}
%\title{Low-Rank Approximation Algorithm for Analysis of Scattering of Multiple Incidents Electromagnetic Wave}

\author{Chung~Hyun~Lee,~\IEEEmembership{Member,~IEEE}, 
Bowoo Jang,~\IEEEmembership{Student Member,~IEEE},
Kyoungil Kwon,~\IEEEmembership{Member,~IEEE}, 
Kyung-Tae Kim,~\IEEEmembership{Member,~IEEE}, and 
Dong-Yeop Na,~\IEEEmembership{Member,~IEEE}% <-this % stops a space
\thanks{
{Chung~Hyun~Lee is with the Division of Semiconductor and Electronics Engineering, Hankuk University of Foreign Studies, Yongin, Republic of Korea, Kyoung Il Kwon is with Agency for Defense Development (Republic of Korea), Daejeon, South Korea, and Bowoo Jang, Kyung-Tae Kim, and Dong-Yeop Na are with the Department of Electrical Engineering, Pohang University of Science and Technology, Pohang, Gyeongsangbuk-do 37673, South Korea (e-mail: chunghyun.lee@hufs.ac.kr; bowoo@postech.ac.kr; kikwon@add.re.kr; kkt@postech.ac.kr; dyna22@postech.ac.kr)}
}
}


\markboth{Preprint}%
{Shell \MakeLowercase{{et al.}}: Bare Demo of IEEEtran.cls for IEEE Journals}

\maketitle

\begin{abstract}
Accurate prediction of the radar cross section (RCS) of chaff clouds requires careful consideration of aerodynamic effects, 
as the orientation and spatial distribution of individual chaff elements evolve significantly after deployment. 
Building upon conventional six-degree-of-freedom (6-DoF) formulations for chaff aerodynamic analysis~\cite{seo2012dynamic}---which assumed straight or two-dimensionally bent geometries---we extend the framework to incorporate arbitrarily curved three-dimensional chaff geometries. 
This extension enables accurate modeling of both flattened and helical dynamics induced by aerodynamic moments acting along the roll, pitch, and yaw directions, thereby providing a more comprehensive and realistic description of chaff motion. 
We then finally develop a coupled aerodynamic--electromagnetic framework that integrates the extended aerodynamic model with our recently developed fast method-of-moments solver~\cite{lee2024fast}, 
which is optimized for efficiently estimating the RCS of million-scale chaff clouds. 
The proposed multiphysics coupled framework allows real-time, first-principles prediction of the monostatic and bistatic RCS of large-scale chaff clouds with arbitrary geometries, orientations, and lengths, accurately incorporating their time-varying aerodynamic evolution. 
Simulation results confirm that the monostatic RCS is strongly influenced by aerodynamic effects, with the coexistence of flattened and helical motions playing a critical role in determining the overall scattering response. 
The proposed framework thus provides a physically grounded and computationally efficient approach for predicting the RCS of large-scale chaff clouds. 
Furthermore, it can be directly extended to radar signal processing applications by utilizing multi-frequency complex-valued far-field responses, 
thereby enabling the reconstruction of Range--Doppler, Range--Angle, and Doppler--Angle maps.
\end{abstract}

\begin{IEEEkeywords}
Chaff, radar cross section, aerodynamics, six-degree-of-freedom equations of motion, electromagnetic scattering, thin-wire approximation, electric-field integral equation, method of moments, sparsification via neglecting far-field coupling, helical motion.
\end{IEEEkeywords}

\IEEEpeerreviewmaketitle

% Introduction

\section{Introduction}
\label{sec:introduction}
\IEEEPARstart{C}{haff} is a widely used radiofrequency (RF) countermeasure deployed by military aircraft, ships, and ground vehicles to disrupt adversary radar systems \cite{wilson1989ewfundamentals}.
Typical chaff elements are slender metallic fibers with a high aspect ratio, usually designed so that their length is approximately half the wavelength of the operational RF band, thereby maximizing their monostatic radar cross section (RCS).
In practice, millions of such fragments are released into the atmosphere, generating a large monostatic RCS that effectively deceives radar sensors.
After a chaff cloud is deployed and dispersed by an explosive charge, the orientation and velocity of individual chaff elements vary significantly due to aerodynamic behavior. 
Since the monostatic RCS is highly sensitive to the orientation and spatial distribution of the chaff cloud \cite{marcus2015bistatic,Seo2010-fo,lee2024fast}, its accurate prediction requires proper consideration of the aerodynamic behavior of chaff particles.

During free fall through air, each chaff element is primarily subjected to gravity and aerodynamic forces. 
It is well known that a perfectly straight and rigid chaff element maintains its orientation during descent due to the cancellation of aerodynamic moments resulting from geometric symmetry~\cite{brunk1975chaff,marcus2004dynamics,nam2009simulations,seo2012dynamic}. 
In practice, however, chaff elements are not perfectly straight but exhibit irregular three-dimensional geometries, arising from manufacturing imperfections and the initial explosive deployment process. 
In particular, during practical manufacturing processes, chaff fibers are rarely produced as perfectly straight elements; mechanical stresses during electroplating, chopping, and post-treatment inevitably induce bending or twisting. 
As described in~\cite{uspat4976828}, even high-strength metal-coated carbon fibers are explicitly tested for their resistance to flaking under twisted or knotted conditions, implying that such geometrical deformation commonly occurs. 
Furthermore, the 98\% undamaged ejection requirement \cite{globalsecurity_chaff} ensures mechanical integrity but does not necessarily guarantee strict linearity of every chaff element, as mild curvature or twist may remain within the tolerance for undamaged.
As a result, arbitrarily bent chaff elements are subjected to geometry-induced aerodynamic moments and damping effects, which typically give rise to flattened orientations and helical trajectories~\cite{brunk1975chaff,marcus2004dynamics}.
Therefore, accurate real-time estimation of the RCS of chaff clouds during free fall necessitates full-wave electromagnetic (EM) modeling that incorporates the aerodynamic responses associated with arbitrarily curved three-dimensional chaff geometries.

The aerodynamic behavior of chaff can be described using six-degree-of-freedom (6-DoF) equations of motion, from first principles.  
Previous studies have demonstrated that the dispersion characteristics of chaff clouds vary markedly depending on the shape and initial conditions of chaff elements, including their orientation, angular velocity, and translational velocity \cite{brunk1975chaff,nam2009simulations,seo2012dynamic}.
For straight chaff, the absence of geometry-induced moments due to symmetry allows the initial orientation to be largely preserved during descent, leading the cloud to expand nearly linearly in radius with time, forming a spherical shell, while the falling attitude remains close to its initial state. 
In contrast, irregular chaff geometries induce aerodynamic moments, causing the cloud radius to converge after a certain time and the individual elements to exhibit coupled translational and rotational motions that eventually develop into flattened or helical trajectories~\cite{brunk1975chaff,nam2009simulations,seo2012dynamic}---phenomena that cannot be captured when chaff is idealized as straight, rigid bodies in conventional 6-DoF formulations.
To address this limitation, previous studies introduced 6-DoF modeling for two-dimensionally bent chaff~\cite{nam2009simulations,seo2012dynamic}.
In this case, the axial and normal drag forces vary locally, enabling the geometry-induced moments to be accounted for and thereby capturing the flattened attitude. 
However, the helical motion cannot still be modeled with a two-dimensional bent configuration.

In this work, we extend previous 6-DoF formulations~\cite{nam2009simulations,seo2012dynamic} by introducing an arbitrarily curved three-dimensional (3D) chaff geometry parameterized by the twisting amplitude and bending radius. 
For clarity, we focus on a representative twisted--bent configuration, which enables accurate modeling of both flattened orientations and helical motions. 
Building upon this formulation, a coupled aerodynamic--electromagnetic framework is developed to predict, from first principles, the time-varying monostatic and bistatic RCS of million-scale chaff clouds with arbitrary shapes, orientations, and lengths. 
The electromagnetic solver is based on our recently developed thin-wire approximate EFIE method-of-moments formulation with sparsification via neglecting far-field coupling (SNFC), called THEM-S~\cite{lee2024fast}, whose accuracy and efficiency have been verified through numerical benchmarks. 
The aerodynamic module is integrated with THEM-S to achieve real-time coupled modeling of freely falling chaff clouds.
Validation is performed in three stages. 
First, the aerodynamic model is validated by analyzing the terminal velocities and helical trajectories of single chaff elements with different geometries, and by comparing the free-fall behavior of $1,000$ chaff elements with the experimental data reported in~\cite{arnott2004radar}. 
Second, the electromagnetic model is validated by comparing the monostatic RCS of single chaffs with HFSS results and the bistatic RCS of $1,000$ chaffs computed using LU decomposition, multi-level fast multipole method (MLFMM), and SNFC approaches. 
Finally, large-scale simulations involving one million chaff elements demonstrate that aerodynamic effects strongly influence the RCS evolution, with the coexistence of flattened and helical motions identified as the dominant physical mechanism.

It should be noted that only a limited number of previous studies have addressed {first-principles-based simulations} that simultaneously capture both aerodynamic behavior and EM characteristics. 
Most existing works on EM properties, which rely on full-wave techniques such as the MoM~\cite{wickliff1974average,kruger2009modelling,perotoni2010electromagnetic,kashyap2018rcs,lee2022analysis} and the finite-difference time-domain (FDTD)~\cite{pandey2013modeling} methods, have been restricted by their {huge computational costs}. 
To overcome this challenge, many studies have adopted {approximation-based approaches}, including the vector radiative transfer (VRT) method~\cite{zuo2021bistatic,zheng2023rcs}, the generalized effective cloud (GEC) model~\cite{marcus2007electromagnetic,Seo2010-fo}, or other simplified techniques~\cite{osman2024backscatter}. 
In contrast, the aerodynamic behavior of chaff clouds has generally been modeled using {phenomenological or statistical approaches}~\cite{marcus2004dynamics,pinchot2005chaff,zhu2018approach}. 
Although some first-principles aerodynamic formulations based on the 6-DoF formulation exist~\cite{brunk1975chaff,huang2018experimental,zhang2022investigation,kim2023modeling}, they typically assume simple straight geometries, except for a few studies addressing two-dimensionally bent chaffs~\cite{nam2009simulations,seo2012dynamic}.

The novelty of this work lies in the development of a {first-principles-based multiphysics algorithm} that simultaneously models aerodynamic and EM behaviors using fast numerical schemes optimized for large-scale chaff cloud RCS analysis. 
In particular, during free-fall simulations, the proposed model captures not only the flattened motion but also the helical motion of chaff induced by geometric asymmetries and fabrication imperfections, thereby enabling accurate real-time tracking of posture evolution and RCS variation in chaff clouds. 
Specifically, the main contributions of this work are summarized as follows:
\begin{enumerate}
    \item Development and integration of a {6-DoF aerodynamic solver} and a {MoM-based electromagnetic solver}, both constructed from first principles and equipped with specialized {computational acceleration techniques}.
    \item Incorporation of {geometric asymmetry}, which naturally arises during chaff fabrication, into the aerodynamic model, enabling accurate reproduction of the {helical motion} that is subsequently reflected in the {time-varying RCS prediction}.
    \item Development of an {accelerated computational framework} capable of rapidly predicting the RCS of {large-scale chaff clouds} consisting of up to one million elements. 
\end{enumerate}
To the best of our knowledge, the proposed {first-principles-based coupled aerodynamic--electromagnetic solver}, which fully accounts for {arbitrarily curved three-dimensional chaff geometries} in large-scale chaff cloud simulations, has not been reported in the prior literature.

\begin{table}[t]
\centering
\caption{List of frequently used parameters in the manuscript.}
\label{tab:parameters}
\renewcommand{\arraystretch}{1.15}
\begin{tabular}{lp{0.65\columnwidth}}
\hline
\textbf{Symbol} & \textbf{Description} \\
\hline
$N_c$ & Number of chaff elements \\
$L$, $L^{(i)}$ & Length of the ($i$-th) chaff element \\
$D$, $D^{(i)}$ & Diameter of the ($i$-th) chaff element \\
$m$, $m^{(i)}$ & Mass of the ($i$-th) chaff element \\
$\rho_{\text{air}}$ & Air density \\
$\mu$ & Dynamic viscosity of air \\
$N_s$ & Number of segments per chaff \\
$N_v=N_s+1$ & Number of total vertices per chaff \\
$N_{iv}=N_v-2$ & Number of internal vertices per chaff \\
$R_b$, $R_b^{(i)}$ & Bending radius of the ($i$-th) chaff \\
$A_t$, $A_t^{(i)}$ & Twisting amplitude of the ($i$-th) chaff \\
$\mathbf{r}_{n}^{(i,\text{mid})}$ & Midpoint of the $n$-th segment of the $i$-th chaff \\
$C_{\parallel}$, $C_{\perp}$ & Drag coefficients in axial and normal directions \\
$Re_{\parallel,n}^{(i)}$, $Re_{\perp,n}^{(i)}$ & Reynolds numbers in axial and normal directions for the $n$-th segment of the $i$-th chaff \\
$g$ & Gravitational acceleration \\
$N_d$ & Number of degrees of freedom for current density modeling \\
$\overline{\mathbf{Z}}$ & Impedance matrix \\
$\mathbf{j}$ & Electric current degrees-of-freedom vector \\
$\mathbf{v}$ & Voltage force vector \\
$\Lambda$ & Rooftop basis function \\
$G$ & Scalar Green’s function in free space \\
$W_T$ & Terminal velocity \\
$d_c$ & Coupling threshold distance for SNFC \\
$d_{\text{mean}}$ & Average inter-element spacing \\
$\lambda$ & Wavelength \\
\hline
\end{tabular}
\end{table}

The manuscript is organized as follows. 
Section~\ref{sec2} describes the aerodynamic module and the six-degree-of-freedom (6-DoF) formulation in detail, 
including the governing equations, chaff geometry modeling, and external force and moment representations. 
Section~\ref{sec3} presents the electromagnetic module based on the THEM-S framework and explains how it enables fast RCS estimation for million-scale chaff clouds. 
Section~\ref{sec4} discusses the integration of the aerodynamic and electromagnetic modules and the multi-threaded CPU implementation. 
Section~\ref{sec5} provides three numerical examples that validate the aerodynamic module. 
Section~\ref{sec6} presents two numerical examples that validate the electromagnetic module. 
Section~\ref{sec7} introduces a large-scale numerical experiment for real-time RCS estimation of a freely falling chaff cloud consisting of one million chaff elements with different geometries. 
Finally, Section~\ref{sec8} provides concluding remarks and outlines potential directions for future research.

To improve readability, several key parameters frequently used throughout the manuscript are summarized in Table~\ref{tab:parameters}.
Here, we adopt the IEEE spherical coordinate convention~\cite{IEEE145-2013}, 
in which the azimuth angle $\phi$ is measured in the $x$--$y$ plane from the $x$-axis, 
and the zenith angle $\theta$ is measured from the $z$-axis takving values from $0^{\circ}$ to $180^\circ$. 
On the other hand, the elevation angle $\varepsilon=90^\circ - \theta$ is defined from the $x$--$y$ plane, taking values from $-90^{\circ}$ to $90^{\circ}$.
In Euler angles, roll, pitch, and yaw are denoted by $\Phi$, $\Theta$, and $\Phi$, respectively.
The time convention $e^{+j\omega t}$ is adopted throughout the entire manuscript.

\section{Aerodynamic module}\label{sec2}
To model the aerodynamic behavior of each chaff element from first principles, we consider the 6-DoF equations of motion \cite{brunk1975chaff,nam2009simulations,seo2012dynamic,sinha2017advanced}.  
The governing equations for the translational velocity of the center of mass (CoM), the angular velocity, and the Euler angles of the $i$-th chaff in the body-fixed frame are given by
\begin{flalign}
\dfrac{d\mathbf{V}^{(i)}}{dt} &= \dfrac{1}{m_i} \mathbf{F}^{(i)}_{\text{ext}} - \bm{\Omega}^{(i)} \times \mathbf{V}^{(i)}, \\
\dfrac{d\bm{\Omega}^{(i)}}{dt} &= \left(\overline{\mathbf{I}}^{(i)}\right)^{-1} \!\cdot\!
\left[ \mathbf{M}^{(i)}_{\text{ext}} - \bm{\Omega}^{(i)} \times \left(\overline{\mathbf{I}}^{(i)} \!\cdot\! \bm{\Omega}^{(i)}\right) \right], \\
\dfrac{d\bm{\eta}^{(i)}}{dt} &= \overline{\mathbf{A}}^{(i)} \cdot \bm{\Omega}^{(i)},
\end{flalign}
where the subscript $i$ denotes the chaff index, $m^{(i)}$ is the mass, $\mathbf{V}^{(i)} = [u^{(i)}, v^{(i)}, w^{(i)}]^{T}$ is the translational velocity, $\bm{\Omega}^{(i)} = [p^{(i)}, q^{(i)}, r^{(i)}]^{T}$ is the angular velocity, and $\bm{\eta}^{(i)} = [\Phi^{(i)}, \Theta^{(i)}, \Psi^{(i)}]^{T}$ represents the Euler angles.  
Here, $\mathbf{F}^{(i)}_{\text{ext}}$ and $\mathbf{M}^{(i)}_{\text{ext}}$ are the external force and moment vectors, respectively, and $\overline{\mathbf{I}}^{(i)}$ denotes the inertia tensor in the body-fixed frame.  
The kinematic transformation matrix is given by
\begin{flalign}
\overline{\mathbf{A}}^{(i)} =
\begin{bmatrix}
1 & \sin\Phi^{(i)}  \tan\Theta^{(i)} & \cos\Phi^{(i)} \tan\Theta^{(i)} \\
0 & \cos\Phi^{(i)} & -\sin\Phi^{(i)} \\
0 & \sin\Phi^{(i)} \sec\Theta^{(i)} & \cos\Phi^{(i)} \sec\Theta^{(i)}
\end{bmatrix}.
\end{flalign}
To transform physical quantities from the body-fixed frame to the inertial frame, the rotation matrix $\overline{\mathbf{R}}^{(i)}$, which defines the orientation of the $i$-th chaff, is defined as the product of the individual roll, pitch, and yaw rotation matrices:
\begin{equation}
\overline{\mathbf{R}}^{(i)} = 
\overline{\mathbf{R}}_{z}\!\left(\Psi^{(i)}\right)
\cdot
\overline{\mathbf{R}}_{y}\!\left(\Theta^{(i)}\right)
\cdot
\overline{\mathbf{R}}_{x}\!\left(\Phi^{(i)}\right),
\label{eq:R_product}
\end{equation}
where
\begin{align}
\overline{\mathbf{R}}_{x}\!\left(\Phi^{(i)}\right) &=
\begin{bmatrix}
1 & 0 & 0 \\
0 & \cos\Phi^{(i)} & \sin\Phi^{(i)} \\
0 & -\sin\Phi^{(i)} & \cos\Phi^{(i)}
\end{bmatrix}, \\
\overline{\mathbf{R}}_{y}\!\left(\Theta^{(i)}\right) &=
\begin{bmatrix}
\cos\Theta^{(i)} & 0 & -\sin\Theta^{(i)} \\
0 & 1 & 0 \\
\sin\Theta^{(i)} & 0 & \cos\Theta^{(i)}
\end{bmatrix}, \\
\overline{\mathbf{R}}_{z}\!\left(\Psi^{(i)}\right) &=
\begin{bmatrix}
\cos\Psi^{(i)} & \sin\Psi^{(i)} & 0 \\
-\sin\Psi^{(i)} & \cos\Psi^{(i)} & 0 \\
0 & 0 & 1
\end{bmatrix}.
\end{align}
The above non-linear equations can be numerically integrated using a fourth-order Runge--Kutta scheme \cite{Hairer2006-xg}. 
Note that, to avoid gimbal-lock instability, we employ quaternions instead of Euler angles.

In the following subsections, we describe how an arbitrary three-dimensional chaff geometry can be modeled and how its aerodynamic behavior can be analyzed in detail. 
According to the slender-body theory \cite{adams1953slender} applicable in the low-to-moderate Reynolds number regime, our apporach is following what has been done in \cite{nam2009simulations,seo2012dynamic} to compute the total aerodynamic force and moment acting on a bent chaff by discretizing the chaff into small segments and summing the local forces and moments acting on each segment.

\subsection{Parametrization of a Three-Dimensionally Twisted--Bent Chaff}
\begin{figure}
\centering            
\includegraphics[width=\linewidth]{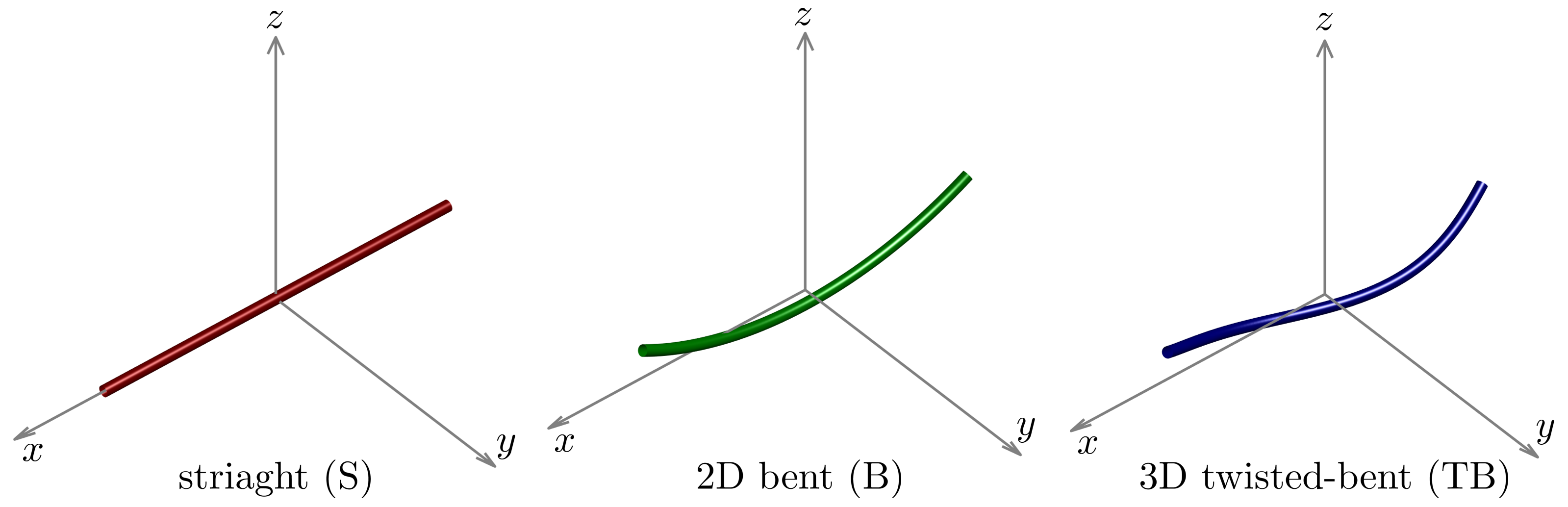}
\caption{Illustration of the three chaff geometries in the body-fixed frame: straight (S), two-dimensionally bent (B), and three-dimensionally twisted--bent (TB).}
\label{fig:chaff_geometry}    
\end{figure}
In the present 6-DoF formulation, the geometry of arbitrarily curved three-dimensional chaff can be modeled through appropriate parameterization.
Here, we consider a twisted--bent geometry of chaff aligned along the $x$-axis in the body-fixed frame, as illustrated in Fig.~\ref{fig:chaff_geometry}. 
For the $i$-th chaff element with total length $L^{(i)}$ and diameter $D^{(i)}$, 
the structure is discretized into $N_s$ straight segments. 
The initial nodal coordinates are parameterized as
\begin{align}
\mathbf{r}_n^{(i,0)} 
&= 
\bigl(
R_b^{(i)} \sin t_n^{(i)},\,
A_t^{(i)} \sin s_n^{(i)},\,
- R_b^{(i)} \cos t_n^{(i)}
\bigr), 
\label{eq:init_coord}
\end{align}
for $n = 1, 2, \ldots, N_v$, where $N_v = N_s + 1$ denotes the total number of nodes per chaff, 
and $R_b^{(i)}$ and $A_t^{(i)}$ represent the bending radius and twisting amplitude, respectively. 
The auxiliary parameters are defined as
\begin{flalign}
t_n^{(i)} &= -\alpha^{(i)} + (n - 1)\Delta \alpha^{(i)},\quad 
s_n^{(i)} = 2\pi \frac{n - 1}{N_s}, 
\nonumber \\
\alpha^{(i)} &= \frac{L^{(i)}}{2 R_b^{(i)}}, 
\quad \Delta \alpha^{(i)} = \frac{2\alpha^{(i)}}{N_s}.
\label{eqn:aux_param}
\end{flalign}
Since the total length of the segmented chaff 
\begin{align}
\ell^{(i)} = \sum_{n=1}^{N_s}
\bigl\|
\mathbf{r}_{n+1}^{(i,0)}  - \mathbf{r}_n^{(i,0)} 
\bigr\|
\label{eq:arc_len}
\end{align}
differs from the prescribed $L^{(i)}$, 
a uniform scaling is applied:
\begin{align}
\mathbf{r}_n^{(i,1)} = 
\frac{L^{(i)}}{\ell^{(i)}}\,\mathbf{r}_n^{(i,0)},
\label{eq:scaled_coord}
\end{align}
for $n = 1,2,\cdots,N_v$.
Finally, the segment midpoints are computed, and the CoM is shifted to the origin to ensure consistency for both aerodynamic and electromagnetic analyses:
\begin{align}
\mathbf{r}_n^{(i,\text{node})} &= \mathbf{r}_n^{(i,1)} - \bar{\mathbf{r}}^{(i,1)}, 
\label{eq:com_shift}
\end{align}
for $n = 1,2,\cdots,N_v$ where $\bar{\mathbf{r}}^{(i,1)} = N_v^{-1}\sum_{n=1}^{N_v}\mathbf{r}_n^{(i,1)}$.
The segment midpoint can be finally expressed by
\begin{flalign}
\mathbf{r}_{n}^{(i,\text{mid})}
=
\sum_{n=1}^{N_s}
\frac{\mathbf{r}_n^{(i)}+\mathbf{r}_{n+1}^{(i)}}{2}
\end{flalign}
for $n = 1,2,\cdots,N_s$.
Note that the twisted--bent chaff configuration naturally reduces to two limiting cases: 
when $A_t^{(i)}\!\to\!0$ and $R_b^{(i)}\!\to\!\infty$, it becomes a straight chaff, 
whereas when $A_t^{(i)}\!\to\!0$ with finite $R_b^{(i)}$, it reduces to a two-dimensionally bent chaff.

\subsection{Moment of Inertia}
Elements of the inertia tensor $\overline{\mathbf{I}}$ can be computed by \cite{sinha2017advanced}
\begin{flalign}
\left[\overline{\mathbf{I}}\right]_{j,k} 
= 
\int \!\left( \sum_{\ell=1}^{3} \zeta_\ell^2 \,\delta_{j,k} - \zeta_j \zeta_k \right) dm,
\label{eq:I_continuous}
\end{flalign}
for $j,k \in \{1,2,3\}$, where $\delta_{j,k}$ denotes the Kronecker delta function, $\zeta_1 = x$, $\zeta_2 = y$, and $\zeta_3 = z$ represent the coordinates of each differential mass element in the body-fixed frame.
For the discretized $i$-th chaff geometry composed of $N_s$ straight segments, Eq.~\eqref{eq:I_continuous} can be approximated by a finite summation as
\begin{flalign}
\left[\overline{\mathbf{I}}^{(i)}\right]_{j,k} \approx \sum_{n=1}^{N_s} 
\left[
\left( \sum_{\ell=1}^{3} \zeta_{\ell,n}^2 \right) \delta_{j,k}
- \zeta_{j,n} \zeta_{k,n}
\right] \Delta m_n^{(i)},
\label{eq:I_discrete}
\end{flalign}
where $(\zeta_{1,n}, \zeta_{2,n}, \zeta_{3,n}) = (x_n^{(i,\text{mid})}, y_n^{(i,\text{mid})}, z_n^{(i,\text{mid})})$ are the coordinates of the segment midpoint, and $\Delta m_{n}^{(i)}=m^{(i)}/N_s$ denotes the mass of the $n$-th segment.  
This discrete representation allows direct computation of the inertia tensor for arbitrarily shaped chaff geometries within the 6-DoF framework.

\subsection{External force and moment}
In the present 6-DoF model, the external forces acting on the $i$-th chaff element consist of the aerodynamic and gravitational forces, which are expressed as follows:
\begin{flalign}
\mathbf{F}_{\text{ext}}^{(i)} &= \mathbf{F}_{\text{aero}}^{(i)} + \mathbf{F}_{\text{gravity}}^{(i)}.
\end{flalign}
Since a twisted-bent chaff experiences different aerodynamic forces in the axial and normal directions locally, for the divided segments, the local forces on each segment are computed, and then integrated to obtain the total force.
The relative velocity of each segment in the body-fixed frame can be represented by considering translational and rotational motions by
\begin{flalign}
\mathbf{V}_{n}^{(i)}
&= 
\mathbf{V}^{(i)} 
- 
\left(\overline{\mathbf{R}}^{(i)}\right)^{T} \cdot \mathbf{v}_{\text{wind}}
+ 
\bm{\Omega}^{(i)} \times \mathbf{r}_{n}^{(i,\text{mid})},
\end{flalign}
where $\mathbf{v}_{\text{wind}}$ represents the background wind speed in the inertial frame.
When decomposing the relative velocity into axial ($\parallel$) and normal ($\perp$) components for each segment, one can compute the total aerodynamic force acting on $i$-th chaff by
\begin{flalign}
\mathbf{F}_{\text{aero}}^{(i)} 
= 
\sum_{n=1}^{N_s}
\mathbf{F}_{\mathrm{aero},n}^{(i)}
\end{flalign}
where
\begin{flalign}
\mathbf{F}_{\mathrm{aero},n}^{(i)}
=
-\frac{\rho_{\text{air}} \Delta A_n}{2}
\sum_{\xi=\left\{\parallel,\perp\right\}}
C_{\xi}(Re_{\xi,n}^{(i)}) \left|\mathbf{V}_{\xi,n}^{(i)}\right|\mathbf{V}_{\xi,n}^{(i)}
\end{flalign}
where $\Delta A_n^{(i)} = L^{(i)} D^{(i)}/N_s$ denotes the projected area, $Re_{\xi,n}^{(i)}$, $C_{\xi}$, and $\mathbf{V}_{\xi,n}^{(i)}$ are the Reynolds number, drag coefficient, and relative velocity component for tangential and normal components of $n$-th segment.
Note that the Reynolds number depending on the relative speed is calculated by \cite{arnott2004radar,nam2009simulations,seo2012dynamic,zhang2022investigation,HUANG20182080}
\begin{flalign}
Re_{\xi,n}^{(i)} = \frac{\rho_{\text{air}}D^{(i)}}{\mu} \left| \mathbf{V}_{\xi,n}^{(i)}\right|
\end{flalign}
for $\xi=\{\parallel,\perp\}$.
Here, following the approach taken in \cite{nam2009simulations,seo2012dynamic}, the normal drag coefficients are calculated using the curve-fit formula proposed by Sucker and Brauer \cite{sucker1975fluid}, which is based on experimental data for the drag ceofficients of a circular cylinder in crossflow at very low Reynolds numbers.
Furthermore, the axial force is evaluated using the analytical solution derived by Curle \cite{curle1980calculation} based on boundary-layer theory.
On the other hand, all segments feel the gravitational force equally; hence it is given by
\begin{flalign}
\mathbf{F}^{(i)}_{\text{gravity}} = \overline{\mathbf{R}}^{(i)} \cdot ( m^{(i)} \mathbf{g} ), 
\quad \mathbf{g} = -g\,\hat{z}.
\end{flalign}
The resulting moment can be obtained as
\begin{flalign}
\mathbf{M}_{\text{ext}}^{(i)}
= \sum_{n=1}^{N_s} \mathbf{r}_{n}^{(i,\text{mid})} \times 
\mathbf{F}_{\mathrm{aero},n}^{(i)}.
\label{eq:moment_sum}
\end{flalign}
Note that the above includes both the geometry-induced moment and the aerodynamic moment-damping effects.
Specifically, the geometry-induced moment arises from the fact that the local aerodynamic forces $\mathbf{F}_{\mathrm{aero},n}^{(i)}$ are not symmetrically distributed with respect to the chaff’s CoM due to its twisted--bent geometry. 
Even when the magnitudes of the local aerodynamic forces are similar, their lever arms $\mathbf{r}_{n}^{(i,\text{mid})}$ vary in direction, producing a net torque that tends to restore or destabilize the chaff orientation. 
At the same time, each $\mathbf{F}_{\mathrm{aero},n}^{(i)}$ contains both the axial and normal aerodynamic force components that depend on the instantaneous local velocity and angle of attack of the segment. 
These components generate a velocity-dependent opposing torque that resists angular motion, which manifests as the aerodynamic moment-damping effect. 
Therefore, by summing the vector products $\mathbf{r}_{n}^{(i,\text{mid})} \times \mathbf{F}_{\mathrm{aero},n}^{(i)}$ over all segments, Eq.~\eqref{eq:moment_sum} simultaneously captures the static torque caused by geometric asymmetry and the dynamic torque due to aerodynamic damping, without requiring separate analytical treatment for each contribution.

\section{Electromagnetic module}\label{sec3}
For the chaff cloud analysis, we recently developed a novel acceleration algorithm optimized for solving the method-of-moments (MoM) linear system derived from a thin-wire approximation of the electric-field integral equation, referred to as THEM-S \cite{lee2024fast}.  
The THEM-S solver accelerates the computation by neglecting weak far-field coupling terms.
Specifically, current-interaction terms in the impedance matrix are truncated when the separation distance between two current segments belonging to different chaffs exceeds a prescribed threshold distance $d_c\approx 2\lambda$.  
This truncation is justified because the EM coupling between different chaff elements is weak owing to electrical insulation, whereas the intra-chaff interactions must be fully retained, as in conventional single-body MoM analyses.
As a consequence of this unique scattering characteristic, when the mean inter-chaff spacing exceeds approximately $2\lambda$, the total coherent scattering of the chaff cloud can be accurately approximated as the incoherent sum of the individual chaff contributions \cite{Garbacz1975,peebles1984bistatic,marcus2015bistatic}.

The core algorithm of THEM-S is summarized as follows.  
Since the diameter $D$ of a chaff element is much smaller than the wavelength (typically $\lambda / D > 100$), the current induced on the chaff surface can be approximated as a one-dimensional current along the wire axis using the thin-wire approximation~\cite{gibson2024method}.  
Accordingly, the current distribution on each chaff can be expressed as 
\begin{flalign}
\mathbf{J}(\mathbf{r}) \approx 
\sum_{i=1}^{N_c}
\sum_{n=1}^{N_{iv}}
j^{(i)}_{n}\Lambda^{(i)}_{n}(\mathbf{r})\hat{\mathbf{t}}^{(i)}_{n}
=
\sum_{p=1}^{N_{d}}
j_{p}
\Lambda_{p}(\mathbf{r})\hat{\mathbf{t}}_p
\label{eqn:current_expression}
\end{flalign} 
where $j$ denotes the DoF of the currents, $N_{iv}$ denotes the number of internal vertices per chaff to be used for modeling current density, $\hat{\mathbf{t}}$ denotes the tangential unit vector, $\Lambda(\mathbf{r})$ denotes the rooftop basis function defined on the internal nodes of each chaff, excluding the two end nodes, since the current vanishes at the chaff tips.
Substituting \eqref{eqn:current_expression} into the redistributed electric-field integral-equation \cite{gibson2024method} and applying the Galerkin testing, one can obtain the following MoM formulation:
\begin{flalign}
\overline{\mathbf{Z}}\cdot \mathbf{j} = \mathbf{v}
\label{TW_EFIE_MOM}
\end{flalign}
where current unknowns are stored in a column vector $\left[\mathbf{j}\right]_p=j_p$, and elements of the impedance matrix $\overline{\mathbf{Z}}=\overline{\mathbf{Z}}^{(1)}+\overline{\mathbf{Z}}^{(2)}$ and the force vector $\mathbf{v}$ can be obtained by 
\begin{flalign}
\left[\mathbf{v}\right]_{q}
&=
-\frac{j}{\omega\mu_0}
\int_{\sigma_q}
\Lambda_{q}(\mathbf{r})
\left[
\hat{\mathbf{t}}_{q}
\cdot
\mathbf{E}_{inc}(\mathbf{r})
\right]
d\mathbf{r},\\
\left[\overline{\mathbf{Z}}^{(1)}\right]_{q,p}
&=
\int_{\sigma_q}
\int_{\sigma_p}
f_{q,p}(\mathbf{r},\mathbf{r}')
d\mathbf{r}'d\mathbf{r},
\\
\left[\overline{\mathbf{Z}}^{(2)}\right]_{q,p}
&=
-\frac{1}{k^2}
\int_{\sigma_q}
\int_{\sigma_p}
h_{q,p}(\mathbf{r},\mathbf{r}')
d\mathbf{r}'d\mathbf{r},
\end{flalign}
for $q,p=1,2,\cdots,N_{d}$ where $\omega$, $k$, and $\mu_0$ represent angular frequency, wavenumber, and permeability, respectively.
Here, $\sigma_q$ denotes a support of $q$\textsuperscript{th} basis function, and
\begin{flalign}
f_{q,p}(\mathbf{r},\mathbf{r}') &=\left[\hat{\mathbf{t}}_{q}\cdot\hat{\mathbf{t}}_{p}\right]\Lambda_{q}(\mathbf{r})\Lambda_{p}(\mathbf{r}')G(\mathbf{r},\mathbf{r}'),\\
h_{q,p}(\mathbf{r},\mathbf{r}') &= \left[\hat{\mathbf{t}}_{q}\cdot\nabla\Lambda_{q}(\mathbf{r})\right]\left[\hat{\mathbf{t}}_{p}\cdot\nabla'\Lambda_{p}(\mathbf{r}')\right]G(\mathbf{r},\mathbf{r}'),
\end{flalign}
where $G(\mathbf{r},\mathbf{r}')=e^{-jk |\mathbf{r}-\mathbf{r}'|}/(4\pi |\mathbf{r}-\mathbf{r}'|)$ denotes the free space scalar Green's function.
In building the impedance matrix, the singularity extraction can be easily done in the analytic fashion while the other off-diagonal elements can be evaluated numerically by using the Gaussian quadrature\cite{gibson2024method}.

In order to solve \eqref{TW_EFIE_MOM} much faster, rather than accounting for all interactions for chaffs, we consider a few near-field coupling terms, resulting in a block banded (sparse) impedance matrix:
\begin{equation}
    \left[\overline{\mathbf{Z}}_{SNFC}\right]_{q,p}
    =
    \begin{cases}
    \left[\overline{\mathbf{Z}}\right]_{q,p}, & d(\Omega_i,\Omega_{j}) \leq d_c\\
    0, & \text{otherwise},
    \end{cases}
\label{equation:neardef}
\end{equation}
assuming $q$\textsuperscript{th} and $p$\textsuperscript{th} vertices are included in $i$\textsuperscript{th} and $j$\textsuperscript{th} chaffs, respectively, where $d(\Omega_i, \Omega_{j})$ denotes a distance between centroids of $i$\textsuperscript{th} and $j$\textsuperscript{th} chaffs, $d_c$ is a threshold distance for neglecting far-field coupling terms, and $\Omega_i$ denotes the support of $i$-th chaff element.
The details about the SNFC are discussed in \cite{lee2024fast}.
Thus, we finally solve the following sparse linear system instead of \eqref{TW_EFIE_MOM}
\begin{flalign}
\overline{\mathbf{Z}}_{SNFC}\cdot \mathbf{j} = \mathbf{v}.
\label{eqn:TW_EFIE_MOM_SNFC}
\end{flalign}

\section{Integration of Aerodynamic and Electromagnetic Modules and FORTRAN Implementation with OpenMP Parallelization}\label{sec4}
\begin{figure}[h]
    \centering
    \includegraphics[width=\linewidth]{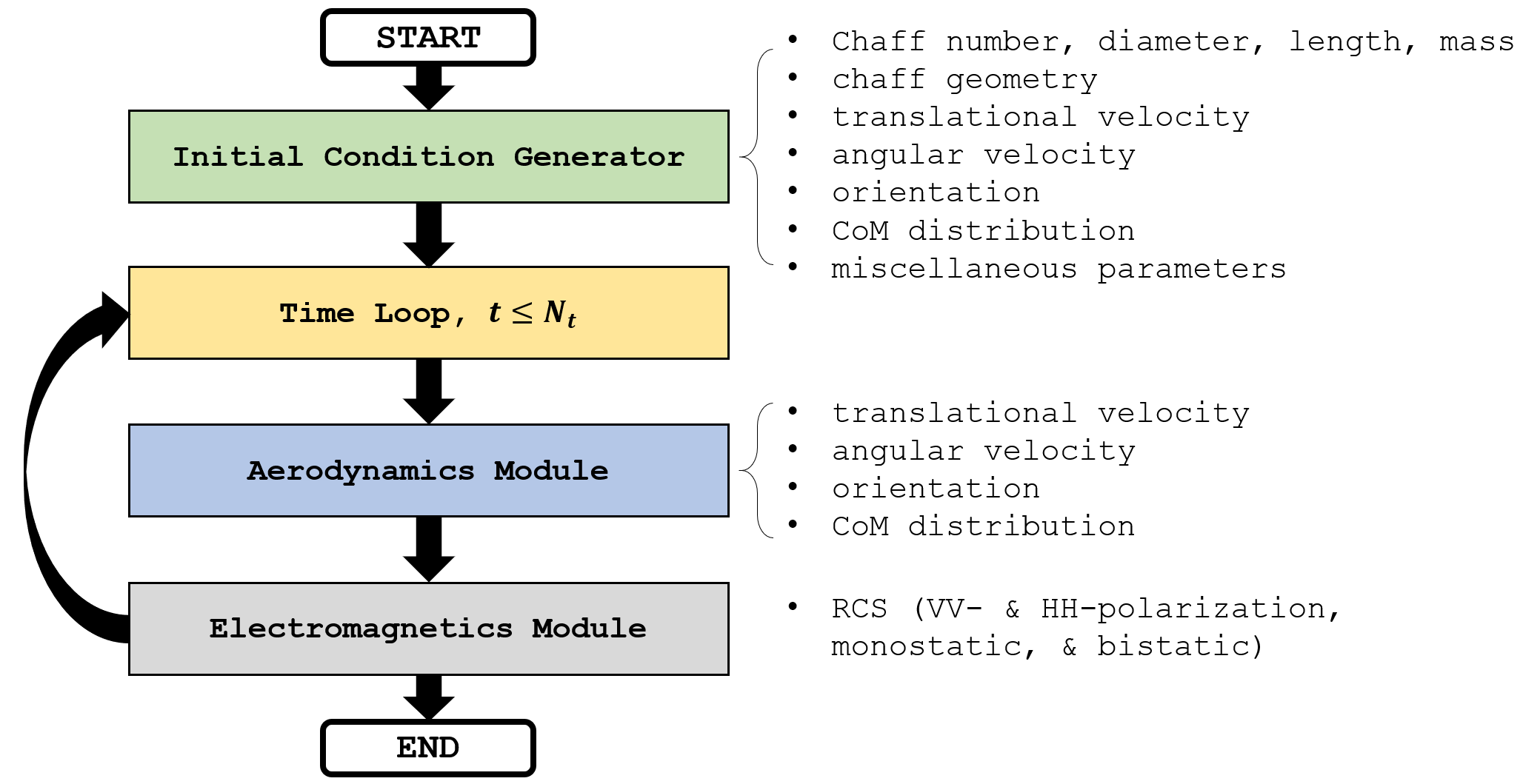}
    \caption{Algorithmic flow of the integrated aerodynamic--electromagnetic modules for estimating the RCS of a chaff cloud while accounting for aerodynamic effects.}
    \label{fig:A_E_module}
\end{figure}

The aerodynamic and electromagnetic modules are integrated within a time-stepping loop to estimate the RCS of a chaff cloud while accounting for aerodynamic effects, as illustrated in Fig.~\ref{fig:A_E_module}. 
Based on the initial conditions of the chaff elements---including the number of chaffs, diameter, length, mass, geometry, translational and angular velocities, orientation, and center-of-mass (CoM) distribution---the time loop begins at $t=0$ and proceeds until $t=N_t$. 
At each time step $\Delta t$, the aerodynamic module updates the translational and angular velocities, orientations, and CoM distributions of all chaff elements. 
Using the updated orientation and positional information, the electromagnetic module computes the RCS of the chaff cloud for various polarizations under both monostatic and bistatic configurations. 
By iteratively coupling these two modules until the end of the simulation, the time-varying RCS of the chaff cloud---evolving in real time due to aerodynamic motion---can be accurately predicted.

In order to accelerate the simulation time of the coupled aerodynamic–electromagnetic solver, we implemented it in a FORTRAN environment with multi-threaded CPU parallelization.
In the aerodynamics module, the Runge–Kutta time-integration scheme for each chaff element is embarrassingly parallelizable, as each trajectory update is independent of the others.
In the electromagnetic module, the most time-consuming step is the construction of the impedance matrix, particularly the evaluation of off-diagonal elements that require twofold numerical quadrature.
To alleviate this computational burden, the impedance matrix assembly was parallelized across threads.
The subsequent linear system solution with a preconditioned iterative solver was also parallelized using OpenMP.
All numerical experiments were conducted on a Linux operating system running on a dual-socket server equipped with Intel Xeon Gold 6430 processors (64 cores) and 1 TB of system memory.

\section{Validation of Aerodynamic Module}\label{sec5}
To validate the aerodynamic module developed in this study, we conducted (i) comparative analyses of the terminal velocity of individual chaff elements as a function of their diameter, (ii) simulations of the free-fall trajectory, velocity, and attitude variations of single chaff elements with different geometries, and (iii) comparisons of the simulation results with experimental data from \cite{arnott2004radar} which hundreds of chaff elements were released under controlled laboratory conditions.  
In the present calculations, external wind and atmospheric turbulence were assumed to be absent.
In what follows, the chaff length $L$ is fixed at $1.78$ [cm].

\subsection{Terminal velocity versus diameter of a single chaff}
\begin{figure}[h]
\centering            
\includegraphics[width=\linewidth]{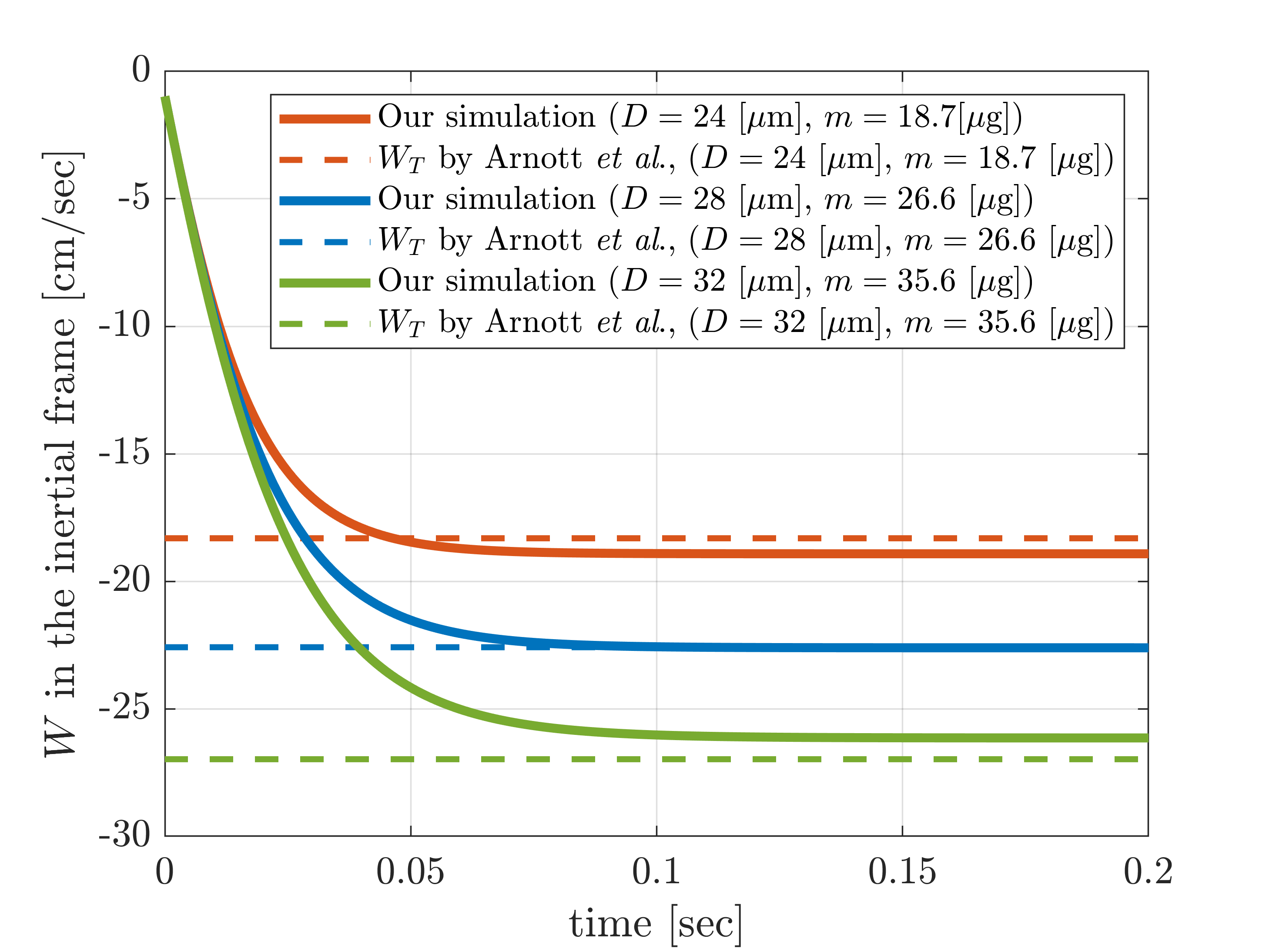}
\caption{Time evolution of the vertical velocity $W$ (z-component of velocity) of a single straight chaff with varying diameters and masses, at a fixed length of 1.78~[cm]. 
The chaff initially falls in a horizontal orientation with zero initial velocity and angular velocity. 
The air density is assumed as $\rho_{\text{air}} = 1.0 \times 10^{-3}\,~[\text{g/cm}^3]$ and the air viscosity as $\mu = 1.83 \times 10^{-4}\,~[\text{g}/(\text{cm}\cdot\text{s})]$, consistent with the conditions in~\cite{arnott2004radar}. 
Here, $W_T$ denotes the terminal velocity.}
\label{fig:SC_W_vs_time}    
\end{figure}

When a straight chaff element initially in a horizontal orientation (elevation = $0^{\circ}$, azimuth = $0^{\circ}$) undergoes free fall, it is accelerated by gravity but soon converges due to aerodynamic drag. 
The terminal velocity of the straight chaff can be expressed in the following semi-empirical form \cite{arnott2004radar}\footnote{
This expression was derived by combining the steady--state form of the equation of motion, $mg = \frac{1}{2}\rho_{\text{air}} A C_D W_T^2$ with the drag correlation for a horizontally oriented infinite cylinder, $C_D = 10.5 R^{-0.63}$ and the Reynolds number definition, $R = W_T D \rho_{\text{air}}\mu^{-1}$.
By expressing the chaff mass and cross--sectional area in terms of its diameter and density, and substituting the above relations into the steady--state equation,
the drag coefficient and Reynolds number can be eliminated.
This leads to a power--law dependence of the terminal velocity on the chaff density, air density, viscosity, and chaff diameter, resulting in the semi--empirical form~\eqref{eqn:Arnott_semi_empirical}.
}:
\begin{flalign}
W_T = 
\frac{
\left[\left(\pi/21\right)\rho_{\text{chaff}} ~ g\right]^{0.73} D^{1.19}
}{
\mu^{0.46}\rho_{\text{air}}^{0.27}
\label{eqn:Arnott_semi_empirical}
}
\end{flalign}
where $\rho_{\text{chaff}}$ is the chaff density, obtained as a function of the chaff diameter $D$; $\rho_{\text{air}}$ denotes the air density (determined by atmospheric temperature and pressure); $g$ is the gravitational acceleration; and $\mu$ is the air viscosity.
For comparison, the values $\rho_{\text{air}}=1.0\times 10^{-3}$ [g/cc] and $\mu = 1.83\times 10^{-4}$ [g/(cm$\cdot$sec)] used in \cite{arnott2004radar} were adopted.
For chaff diameters of $D=24$ [$\mu$m], $D=28$ [$\mu$m], and $D=32$ [$\mu$m], the time-dependent velocity profiles obtained from our 6-DoF solver were compared with the terminal velocities reported in \cite{arnott2004radar} (see Fig.~\ref{fig:SC_W_vs_time}).
The correlation between the chaff mass and its diameter is presented in \cite{arnott2004radar}, and the mass generally increases with increasing diameter. 
As the chaff diameter increases, the mass becomes larger, leading to a stronger influence of gravity relative to aerodynamic drag, and hence a higher terminal velocity.
In all cases, velocity convergence was observed within $0.1$~[sec], and our simulation results showed great agreement with the semi-empricial predictions in \cite{arnott2004radar} with the very small discrepancy (less than $4\%$ of the relative error). 
This validates the correctness of the aerodynamic drag modeling for chaff in vertical free fall.

\subsection{Comparison of the free fall of single straight, 2D-bent, and 3D twisted-bent chaff}
\begin{figure*}[t]
    \centering
    \begin{subfigure}{0.4\linewidth}
        \centering
        \includegraphics[width=\linewidth]{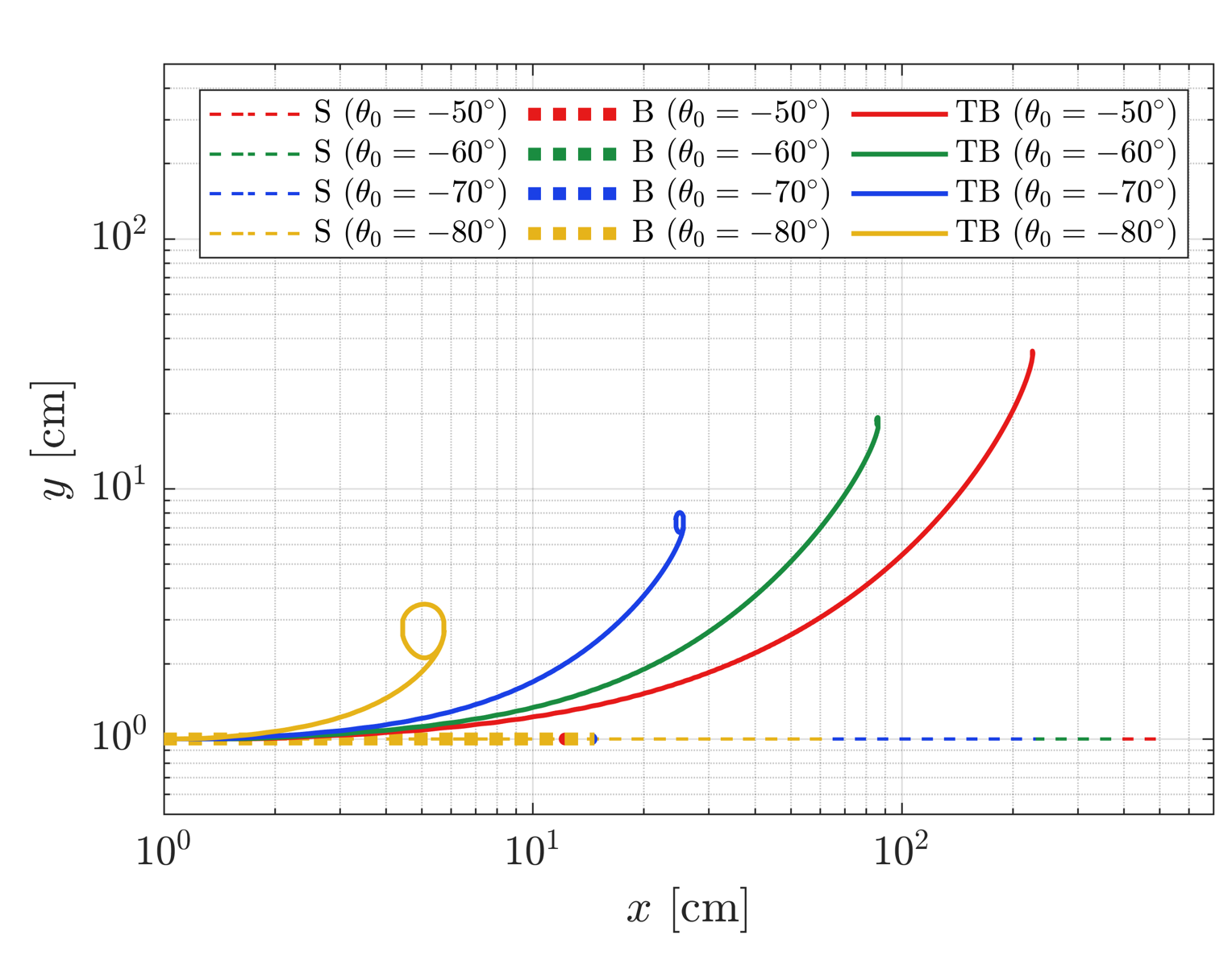}
        \caption{Chaff trajectory on the $xy$ plane.}
        \label{fig:SC_S_B_TB_xy}
    \end{subfigure}
    \quad\quad
    \begin{subfigure}{0.4\linewidth}
        \centering
        \includegraphics[width=\linewidth]{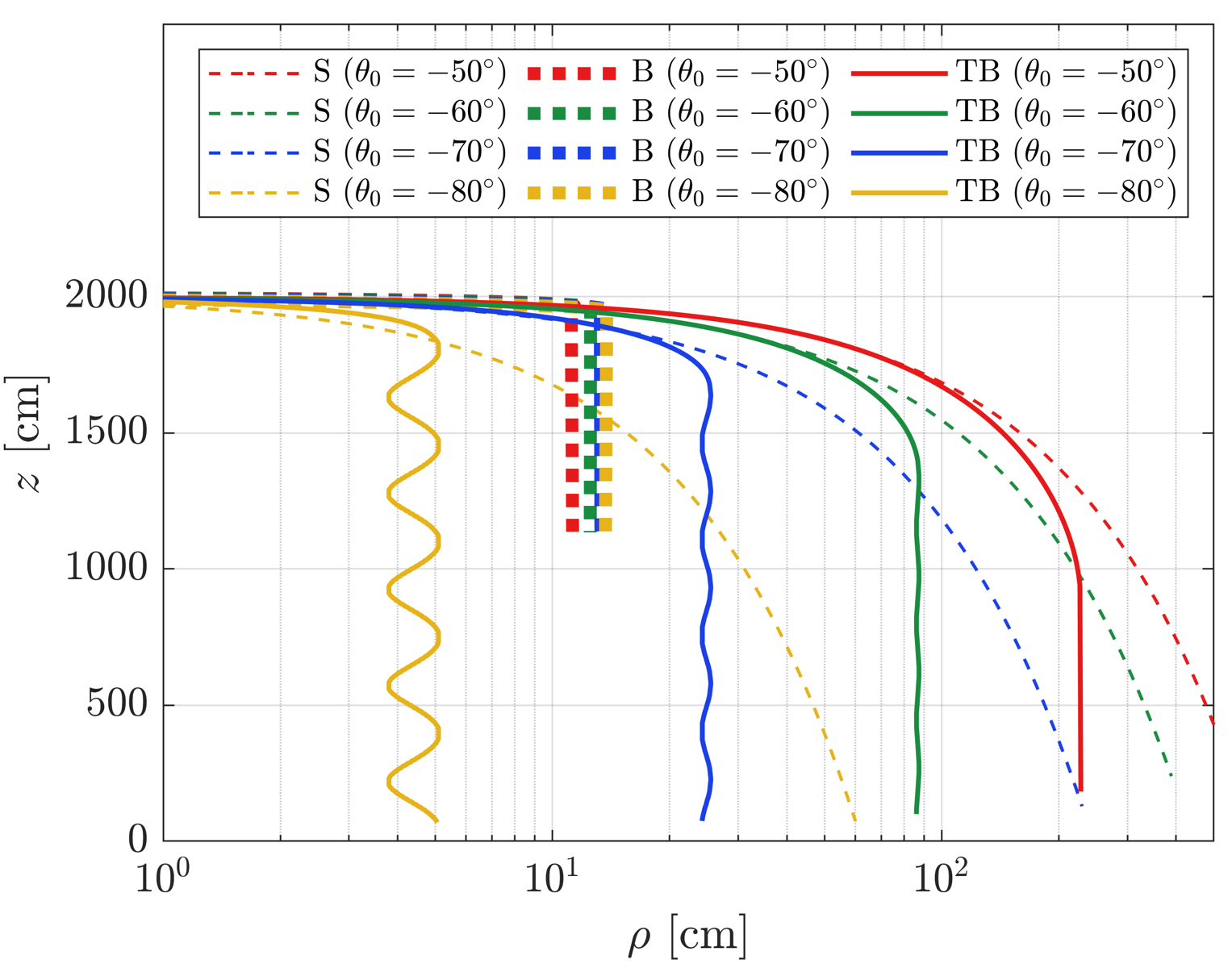}
        \caption{Chaff trajectory on the $\rho$–$z$ plane.}
        \label{fig:SC_S_B_TB_rhoz}
    \end{subfigure}
    \\
    \begin{subfigure}{0.4\linewidth}
        \centering
        \includegraphics[width=\linewidth]{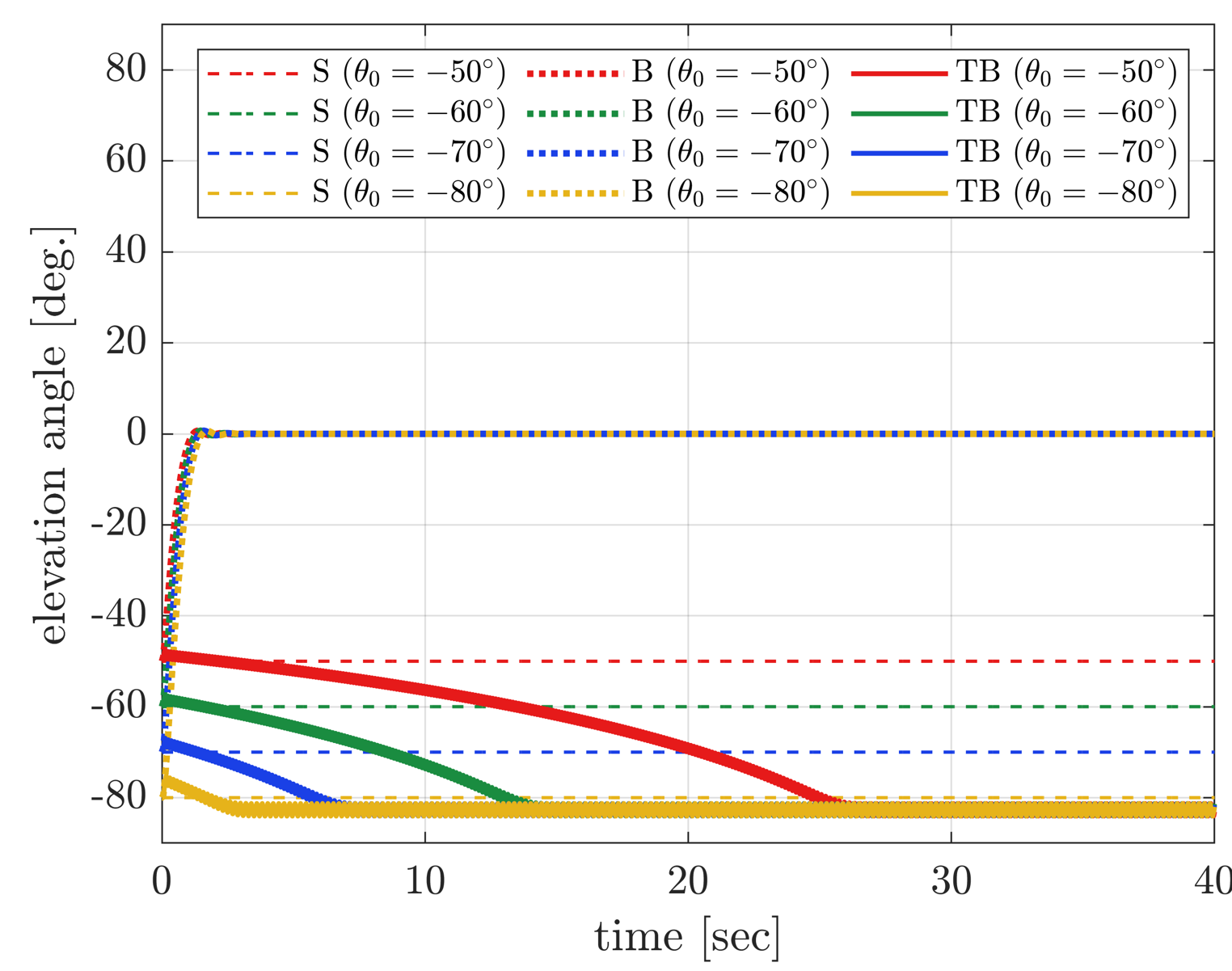}
        \caption{Chaff elevation angle as a function of time.}
        \label{fig:SC_S_B_TB_telev}
    \end{subfigure}
    \quad\quad
    \begin{subfigure}{0.4\linewidth}
        \centering
        \includegraphics[width=\linewidth]{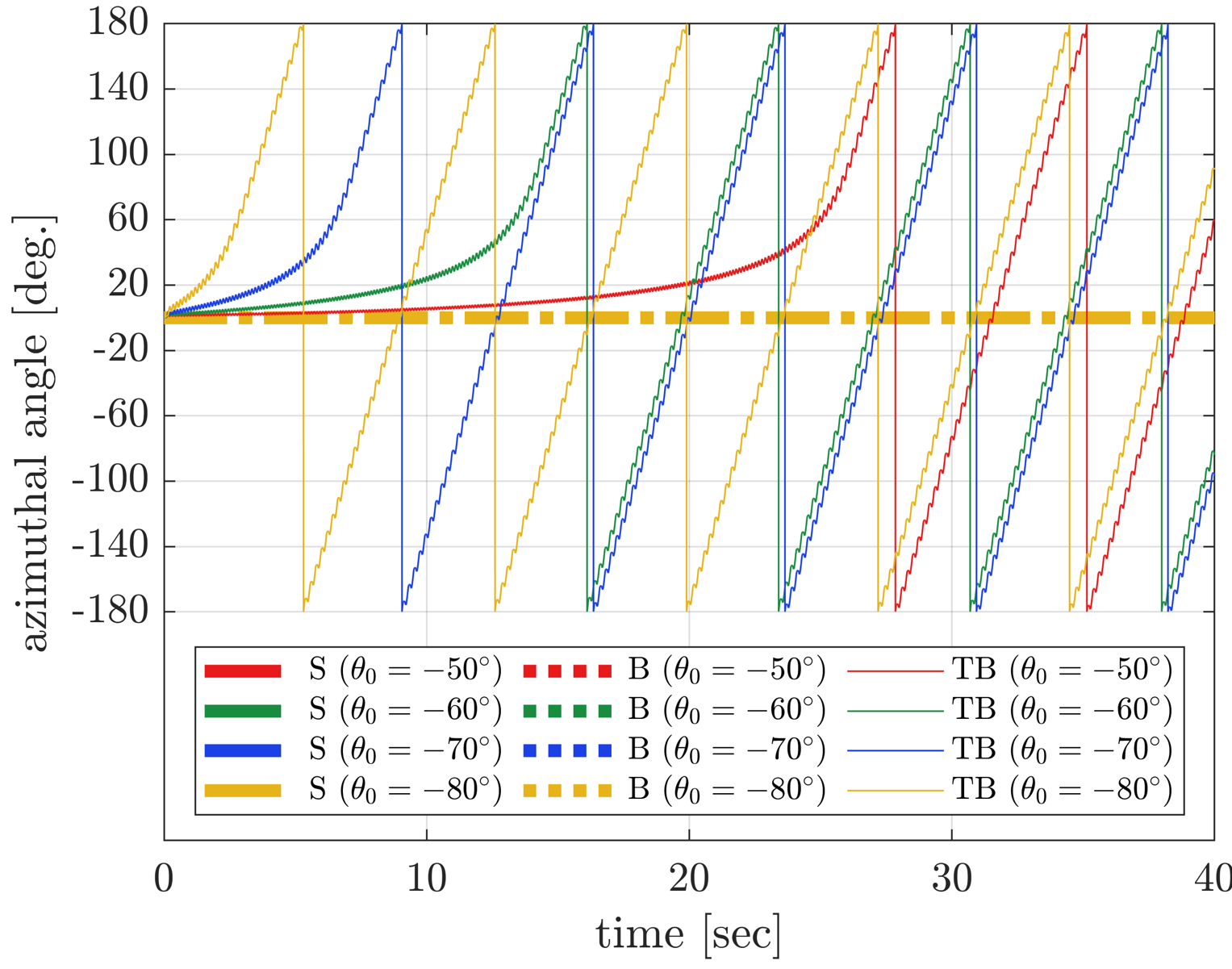}
        \caption{Chaff azimuthal angle as a function of time.}
        \label{fig:SC_S_B_TB_tazim}
    \end{subfigure}
    \caption{Falling characteristics of single chaffs with three different geometries: straight (S), 2D bent (B), and 3D twisted-bent (TB). 
    The S-type chaff maintains its initial orientation with slight lateral slip due to drag anisotropy, the B-type chaff rapidly converges to a horizontal posture within about 2~[sec] due to aerodynamic damping, and the TB-type chaff exhibits helical motion whose onset depends on the initial orientation. 
    These results demonstrate that helical motion arises only with three-dimensional perturbations, highlighting the importance of modeling geometric irregularities for accurate chaff-cloud RCS analysis.}
    \label{fig:SC_geom_falling_results}
\end{figure*}
An ideally straight and perfectly rigid chaff experiences no variation in aerodynamic moment during free fall due to geometric symmetry and therefore maintains its initial orientation throughout the descent.
In reality, slight geometric imperfections such as bending introduce asymmetry, generating moment variations and aerodynamic damping that strongly influence fall dynamics. 
Previous studies~\cite{brunk1975chaff,arnott2004radar,marcus2004dynamics} identified three representative falling behaviors: horizontal convergence, vertical helical motion, and intermediate modes. 
These effects are incorporated into the present 6-DoF simulation, considering three cases: straight, two-dimensionally bent, and three-dimensionally twisted--bent chaff.

Our numerical simulation conditions are as follows. 
Single chaff elements with straight (S), 2D bent (B), and 3D twisted-bent (TB) geometries are released from various initial orientations and allowed to fall freely for 40 [sec]. 
During the fall, we tracked the trajectories in the $xy$-plane and $\rho$–$z$ plane, as well as the time evolution of elevation and azimuth angles, in order to investigate the falling dynamics characteristic of each geometry. 
The simulation parameters were assumed as $m = 27.2 \times 10^{-6}\,~[\text{g}]$, $D = 28.42 \times 10^{-4}\,~[\text{cm}]$, $\rho_{\text{air}} = 1.204 \times 10^{-3}\,~[\text{g/cm}^3]$, $\mu = 1.825 \times 10^{-4}\,~[\text{g/(cm·s)]}$, $R_b = 10L$ (bending radius), $A_t = 5 \times 10^{-3} L$ (twist amplitude), and $N_{s} = 100$ (number of segments). 
The computed results are illustrated in Fig.~\ref{fig:SC_geom_falling_results}.
It can be observed that the S-type chaff exhibits no variation in aerodynamic moment and therefore maintains its initial orientation throughout the entire fall, as expected. 
Due to the difference between horizontal and vertical drag forces, its trajectory appears to slip along the $x$-axis. 
In the case of the B-type chaff, a pitch moment arises from its geometry; however, aerodynamic damping causes it to converge rapidly into a horizontal orientation within approximately 2~[sec]. 
Before this convergence, the trajectory displays displacement along the $x$-axis, but once the chaff has settled into the horizontal orientation, it experiences significantly larger vertical drag, resulting in a reduced terminal velocity and the disappearance of $x$-axis displacement. Interestingly, for the TB-type chaff, depending on the initial orientation, the posture gradually converges to a vertical configuration over time. 
Subsequently, the trajectory in both the $xy$ plane and the $\rho$–$z$ plane, as well as the azimuth angle, evolves almost linearly with time, thereby exhibiting a helical motion. 
Here, $\rho = \sqrt{x^2 + y^2}$ is defined as the radial distance from the $z$-axis in cylindrical coordinates.
It is also observed that as the initial orientation approaches a more horizontal posture, the time required to transition into the helical motion becomes longer. 
In other words, the helical motion can only occur when the chaff geometry incorporates three-dimensional perturbations. 
Since the RCS in both VV and HH polarizations strongly depends on the chaff orientation, accurate aerodynamic modeling that accounts for such three-dimensional geometric perturbations is crucial for reliable prediction of chaff-cloud RCS.

\subsection{Free fall of multiple chaffs}\label{sec:aero_multiple_chaff}
\begin{figure}[h]
\centering            
\includegraphics[width=\linewidth]{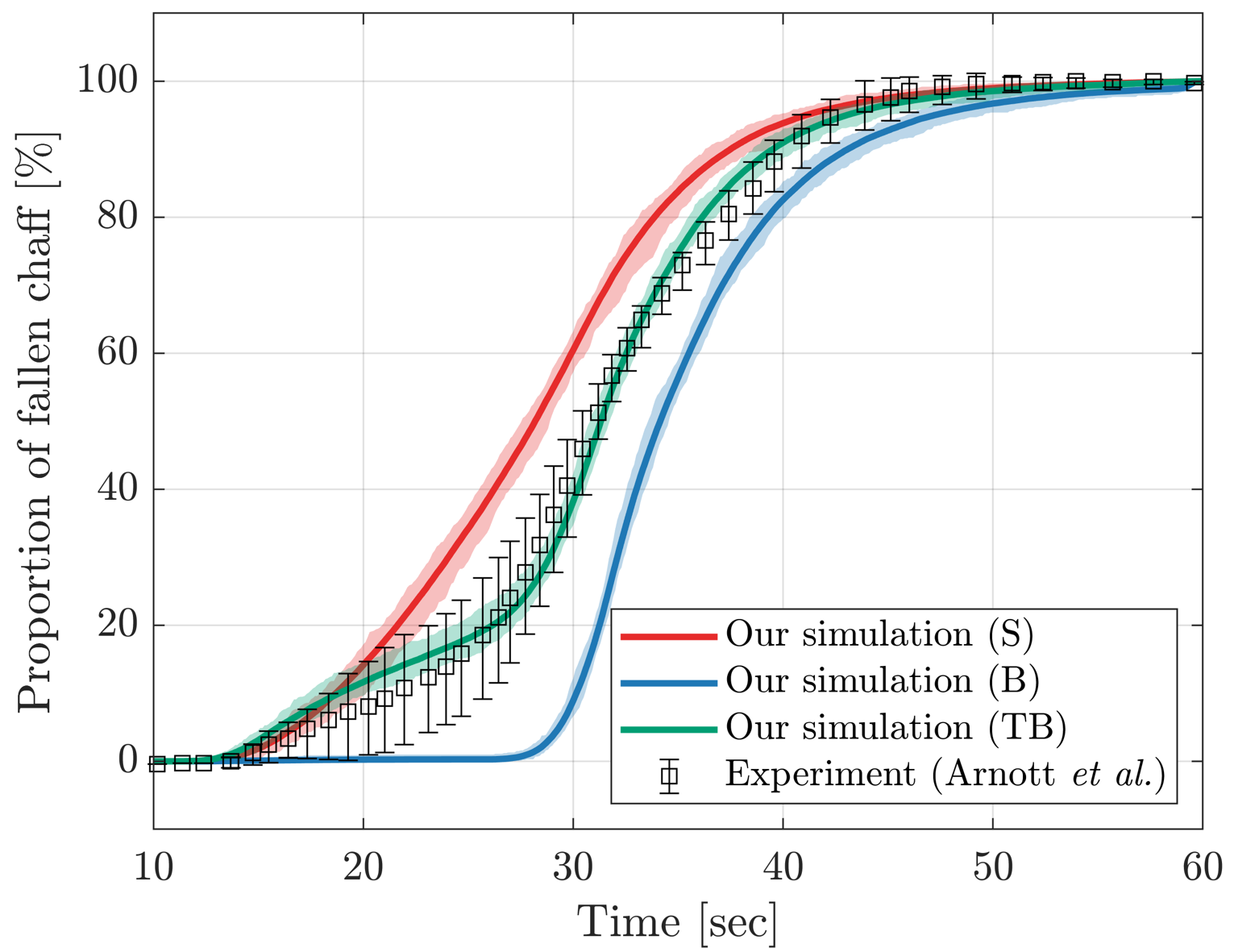}
\caption{Comparison of the fraction of fallen chaff over time between experiments~\cite{arnott2004radar} and numerical simulations for straight (S), bent (B), and twisted-bent (TB) chaff geometries. Solid lines represent the mean of 100 Monte Carlo trials, shaded regions indicate upper and lower bounds, and black square markers with error bars denote experimental data.}
\label{fig:Aero_Exp_Comp}    
\end{figure}

\begin{figure}[h]
\centering            
\includegraphics[width=\linewidth]{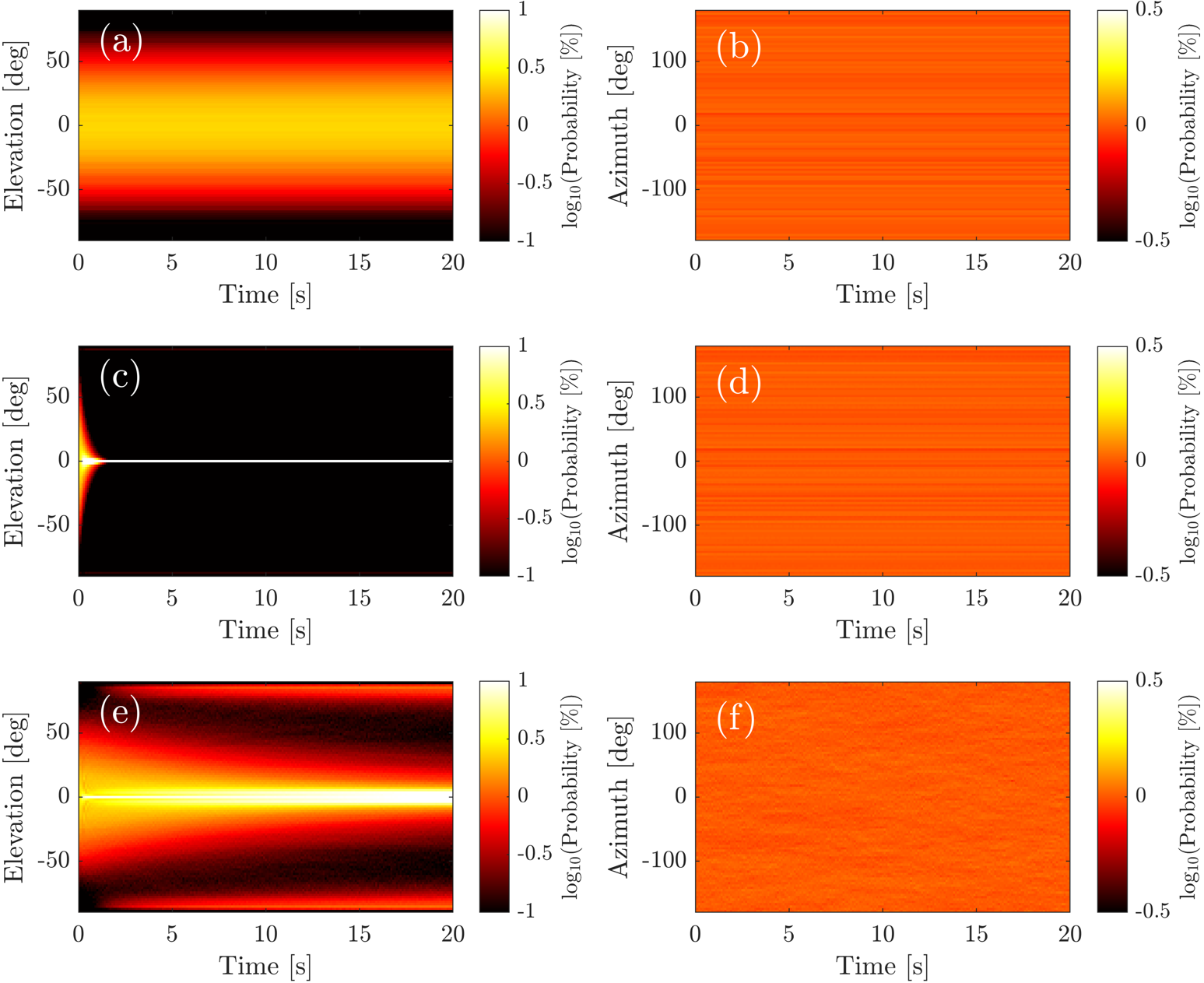}
\caption{Time evolution of the elevation and azimuth angles during the first 20~[sec] for straight (S), bent (B), and twisted-bent (TB) chaff. The S-chaff maintains its initial orientation, the B-chaff rapidly converges to a horizontal posture due to aerodynamic damping, and the TB-chaff exhibits helical motion with nearly uniform azimuthal distribution.}
\label{fig:Aero_Exp_Elev_Azim}    
\end{figure}

To validate the aerodynamic module developed in this study, we compared our numerical results with experimental data obtained from laboratory free-fall tests of several hundred chaff particles~\cite{arnott2004radar}. 
In the experiment, 200 chaffs were released from an initial height of $z_0 = 802$~[cm], and the fraction of collected chaff mass relative to the total released mass was measured. 
Benchmarking this setup, our numerical simulations considered 1,000 chaffs with straight (S), bent (B), and twisted-bent (TB) geometries, also released from $z_0 = 802$~[cm]. 
For each geometry, we performed 100 independent trials with randomized initial conditions generated via a Monte Carlo approach, and statistically analyzed the results.  
Note that in this study, both the initial translational and angular velocities are assumed to be zero. 
The initial orientations of the chaff elements are treated as random variables. 
Specifically, the azimuthal angle follows a uniform distribution between $0^{\circ}$ and $360^{\circ}$, 
while the elevation angle follows a normal distribution with a mean of $0^{\circ}$ and a standard deviation of $30^{\circ}$. 
These initial conditions are benchmarked against the experimental setup described in~\cite{arnott2004radar}, in which the chaffs were initially placed on a metallic plate that was swung downward to release them into free fall.
In the simulation, the physical and aerodynamic properties of each chaff element were assigned as follows. 
The diameter of the chaffs was modeled as a random variable following a normal distribution with a mean of $27$~[$\mu$m] and a standard deviation of $3$~[$\mu$m]. 
The mass of each chaff was determined based on the experimentally measured diameter–mass distribution curve reported in~\cite{arnott2004radar}. 
All chaffs were assumed to have an identical length of $1.78$~[cm].
The bending radius $R_b$ of each chaff was defined as
\begin{equation}
R_b = L + \mathrm{R.V.}(10L),
\end{equation}
where $\mathrm{R.V.}(10L)$ is a random variable uniformly distributed between $0$ and $10L$. 
Similarly, the twist amplitude $A_t$ was defined as
\begin{equation}
A_t = \pm \mathrm{R.V.}(0.1L),
\end{equation}
where $\mathrm{R.V.}(0.1L)$ is uniformly distributed between $0$ and $0.1L$.
The air density and dynamic viscosity were set to $\rho_{\mathrm{air}} = 1.225\times10^{-3}~[\mathrm{g/cm^3}]$ and $\mu = 1.8\times10^{-4}~[\mathrm{g/(cm\cdot s)}]$, respectively.
Fig.~\ref{fig:Aero_Exp_Comp} shows the time evolution of the fraction of fallen chaff for the S-, B-, and TB-type cases, compared with the experimental data~\cite{arnott2004radar}. 
The red, blue, and green solid lines correspond to S-, B-, and TB-shaped chaff, respectively. 
Each solid line represents the mean of 100 trials, while the surrounding semi-transparent bands denote the upper and lower bounds across all trials. 
The experimental data are plotted as black square markers with error bars. 
The results show that S-type chaff tends to fall faster than the experimental measurements, whereas B-type chaff falls more slowly. 
This discrepancy arises because S-type chaff maintains its initial orientation---uniformly distributed between horizontal and vertical---leading to a relatively constant increase in the fallen fraction over time. 
In contrast, B-type chaff quickly converges to a horizontal posture within a few seconds due to aerodynamic damping, which increases vertical drag and reduces its fall speed, resulting in delayed arrival times. 
Interestingly, the TB-type chaff simulation results show excellent agreement with the experimental data. 
Specifically, the simulations successfully reproduce the slow increase in the fallen fraction during 12-29 [sec], the rapid rise between 29-32 [sec], and the subsequent gradual saturation. 
These three stages correspond to nearly vertical helical motion, soft helical motion, and horizontally aligned chaff, respectively.
Fig.~\ref{fig:Aero_Exp_Elev_Azim} further illustrates the evolution of elevation and azimuth angles within the first 20~[sec] for each geometry. 
It can be clearly observed that the S-type chaff retains its initial orientation throughout the fall, the B-type chaff chaff rapidly converges to a horizontal posture within approximately 2.5~[sec], and the TB-type chaff gradually settles into imperfect horizontal or vertical orientations while exhibiting uniformly distributed azimuth angles, thereby producing helical motion.  
Overall, these numerical simulations provide the first demonstration that the free-fall dynamics of real chaff can be almost perfectly reproduced, confirming that helical motion is the dominant mode of descent in practice.  

\section{Validation of Electromagnetic Module}\label{sec6}
In this section, two numerical examples are presented to verify the accuracy and validity of THEM-S in estimating the monostatic and bistatic RCS of chaff clouds with arbitrary three-dimensional geometries.
First, the monostatic RCS of a single chaff element is compared among the S-, B-, and TB-type geometries.
Second, the bistatic RCS of a chaff cloud composed of 1,000 S-, B-, and TB-type chaff elements is evaluated.

\subsection{Monostatic RCS of a Single Chaff with Different Shapes}
\begin{figure}[t]
    \centering
    \begin{subfigure}{\linewidth}
        \centering
        \includegraphics[width=\linewidth]{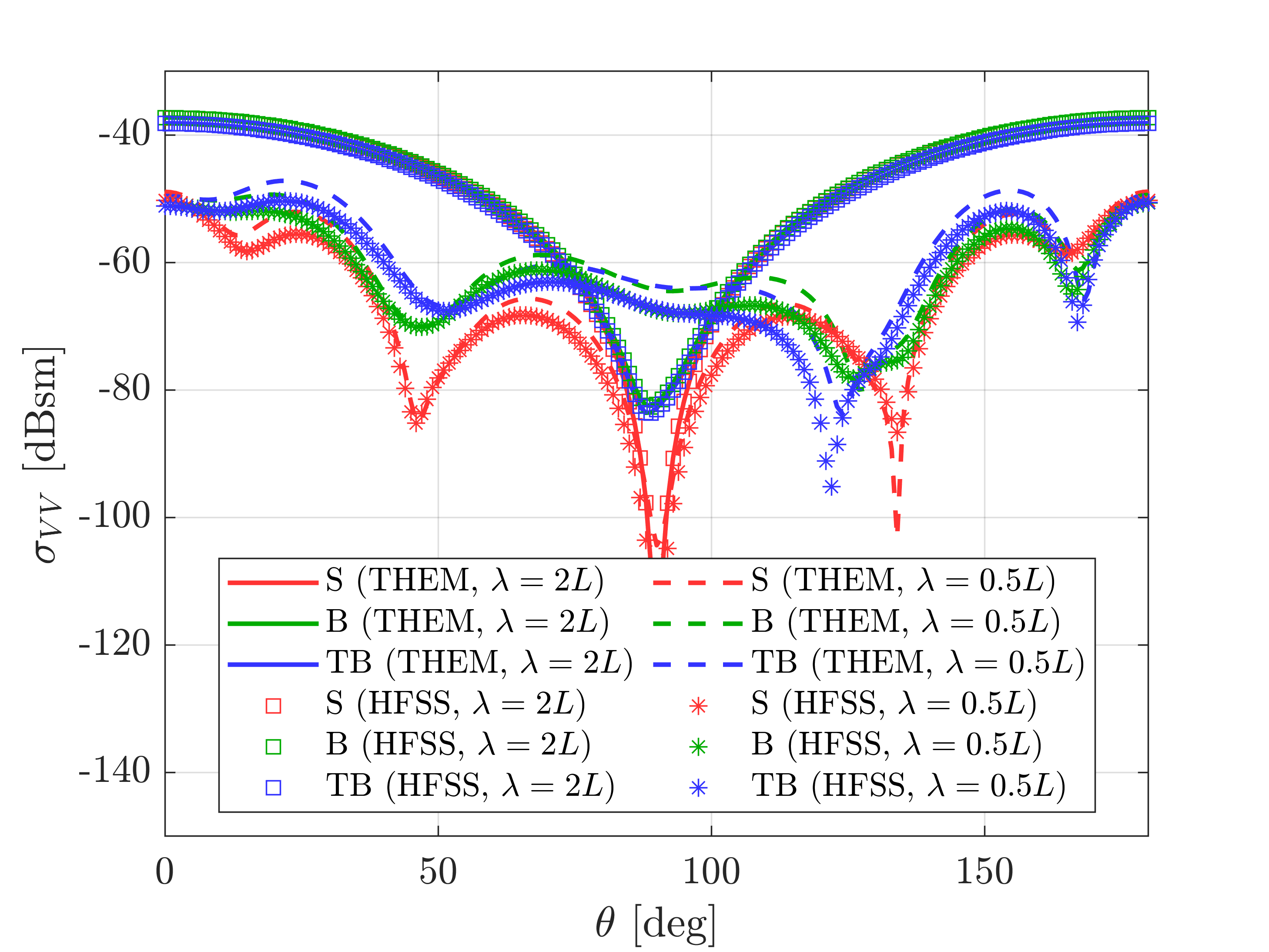}
        \caption{Monostatic VV-polarized RCS.}
        \label{fig:SC_S_B_TB_RCS_VV}
    \end{subfigure}
    \begin{subfigure}{\linewidth}
        \centering
        \includegraphics[width=\linewidth]{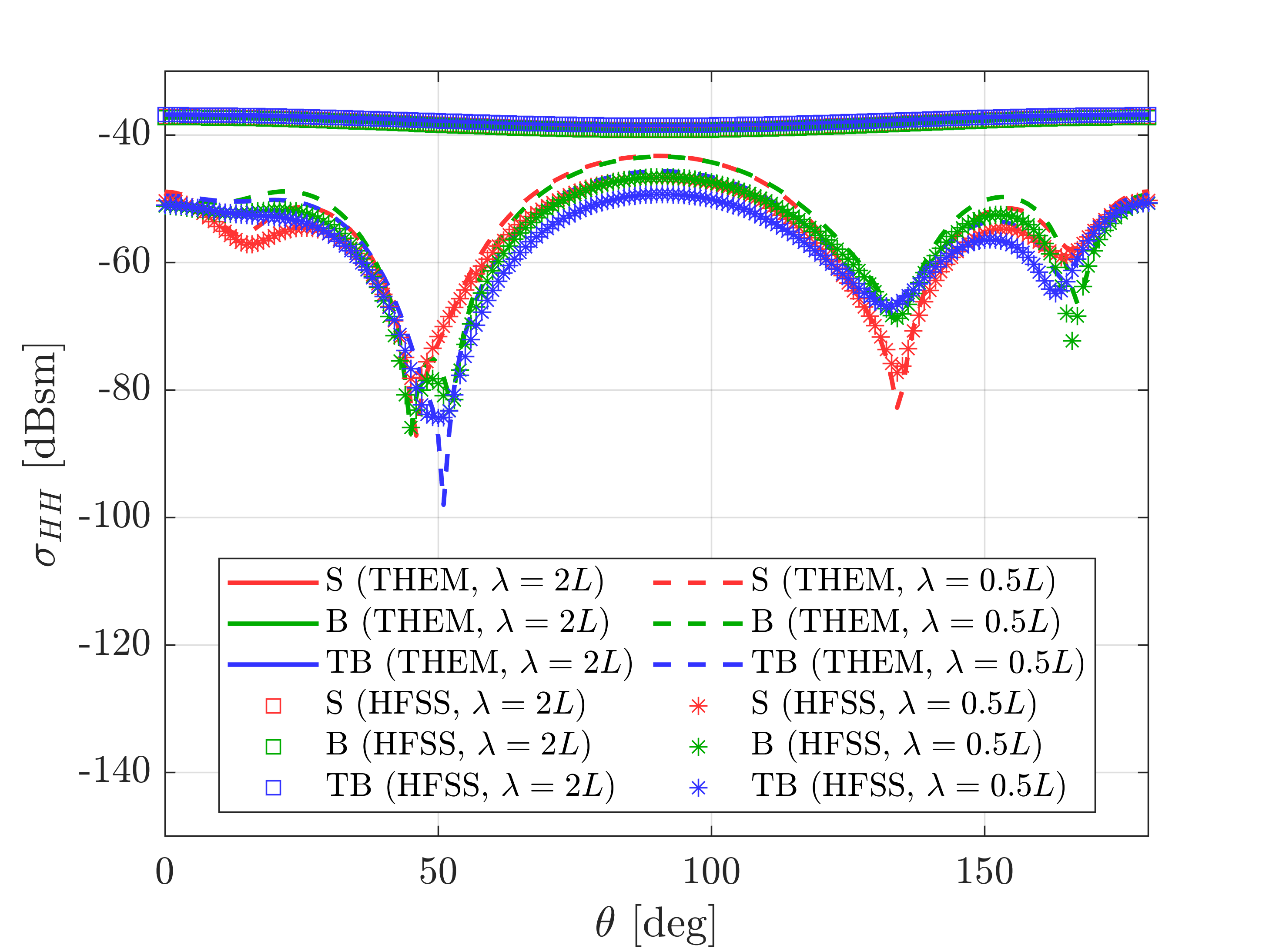}
        \caption{Monostatic HH-polarized RCS.}
        \label{fig:SC_S_B_TB_RCS_HH}
    \end{subfigure}
    \caption{Computed monostatic RCSs of the S-, B-, and TB-type chaff at $\phi_{\mathrm{inc}} = 45^{\circ}$ as a function of the incident zenith angle $\theta_{\mathrm{inc}}$, compared with the results obtained from the HFSS IE solver.}
    \label{fig:SC_mRCS}
\end{figure}

To evaluate how the RCS of a single, three-dimensionally bent chaff considered in the aerodynamic module differs from that of a straight chaff, the RCS values of the S-, B-, and TB-chaff were compared when the chaff length corresponds to half and double the wavelength. 
The results obtained using the proposed THEM computation (without SNFC acceleration) were benchmarked against those from the integral equation (IE) solver in HFSS~\cite{ansysHFSS}. 
Figs.~\ref{fig:SC_S_B_TB_RCS_VV} and \ref{fig:SC_S_B_TB_RCS_HH} show the $\sigma_{VV}$ and $\sigma_{HH}$ results, respectively, plotted as functions of the zenith angle. 
The incident wave is assumed to have an azimuth angle of $\phi_{\mathrm{inc}} = 45^\circ$. 
For modeling the B- and TB-chaff, the parameters $R_b = 2L$~[m] and $A_t = 0.02L$ were used. 
As shown in Figs.~\ref{fig:SC_S_B_TB_RCS_VV} and \ref{fig:SC_S_B_TB_RCS_HH}, when the chaff length corresponds to half a wavelength, the RCS values of the S-, B-, and TB-chaff are nearly identical, in good agreement with the HFSS IE results. 
This indicates that geometric deformation has little effect on the half-wavelength resonance. 
In contrast, when the chaff length corresponds to double the wavelength, the monostatic RCSs of the S-, B-, and TB-chaffs show noticeable differences. 
As the electrical length increases, geometric distortion increasingly affects the current resonance. 
In particular, for $\sigma_{VV}$, the S-chaff exhibits a perfect null at $\theta = 90^\circ$, whereas the B- and TB-chaffs show a relaxation of this null due to shape deformation. 
For $\sigma_{HH}$, the S- and B-chaffs display nearly identical behavior between $\theta = 25^\circ$ and $175^\circ$, while the TB-chaff exhibits slight deviations attributed to three-dimensional bending. 
Compared with the results from the HFSS IE solver, a slight deviation is observed, which arises from the fact that as the wavelength becomes shorter, the electrical length of the chaff diameter increases, and the effect of the diameter on the RCS computation gradually appears. 
Nevertheless, the null positions and overall pattern trends with respect to $\theta$ remain nearly identical, and the discrepancy remains within only a few percent. 
Therefore, the computational accuracy of the proposed THEM approach is still well guaranteed at the higher frequency.

\subsection{Bistatc RCS of 1,000 Chaffs with Different Shapes}
We next consider bistatic RCS calculations for a chaff cloud composed of $N_c = 1,000$ chaff elements with S-, B-, and TB-type geometries. 
Each chaff is discretized into $N_s=10$ segments, corresponding to 9 basis functions per element, resulting in a total of 9,000 basis functions per cloud. 
Here, diameter $D$, $R_b$, $A_t$, initial orientations of chaff elements are random-variabled in the same fashion used in \ref{sec:aero_multiple_chaff}.
The chaff elements are randomly distributed within a spherical cloud of radius $0.5\,~[\text{m}]$, which approximates the realistic deployment condition where individual dipoles disperse in three dimensions without preferred alignment. 
For consistent comparison, all three cloud configurations share the identical random spatial and angular distributions. 
The average inter-element spacing 
\begin{equation}
d_{\text{mean}} = \left(\frac{V_{\text{cloud}}}{N_c}\right)^{1/3},
\end{equation}
was measured to be $80.2\,~[\text{mm}]$. 
The same value of $d_{\text{mean}}$ will also be used later in the subsequent section for the case of a million chaff elements to ensure consistency in cloud density.

\begin{figure*}[t]
  \centering
  % First image
  \begin{minipage}[b]{0.32\textwidth}
    \centering
    \includegraphics[width=\textwidth]{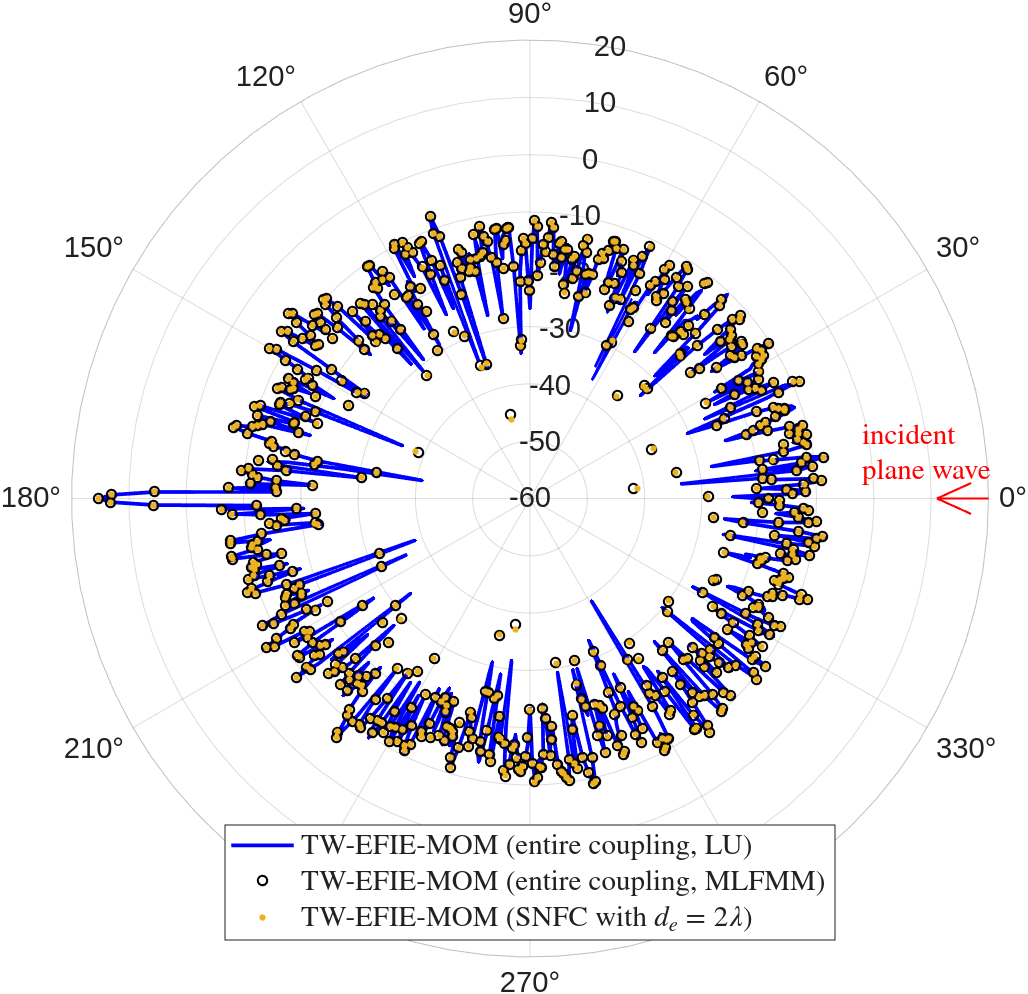}
    \caption*{(a) Straight chaff cloud}
  \end{minipage}
  \hfill
  % Second image
  \begin{minipage}[b]{0.32\textwidth}
    \centering
    \includegraphics[width=\textwidth]{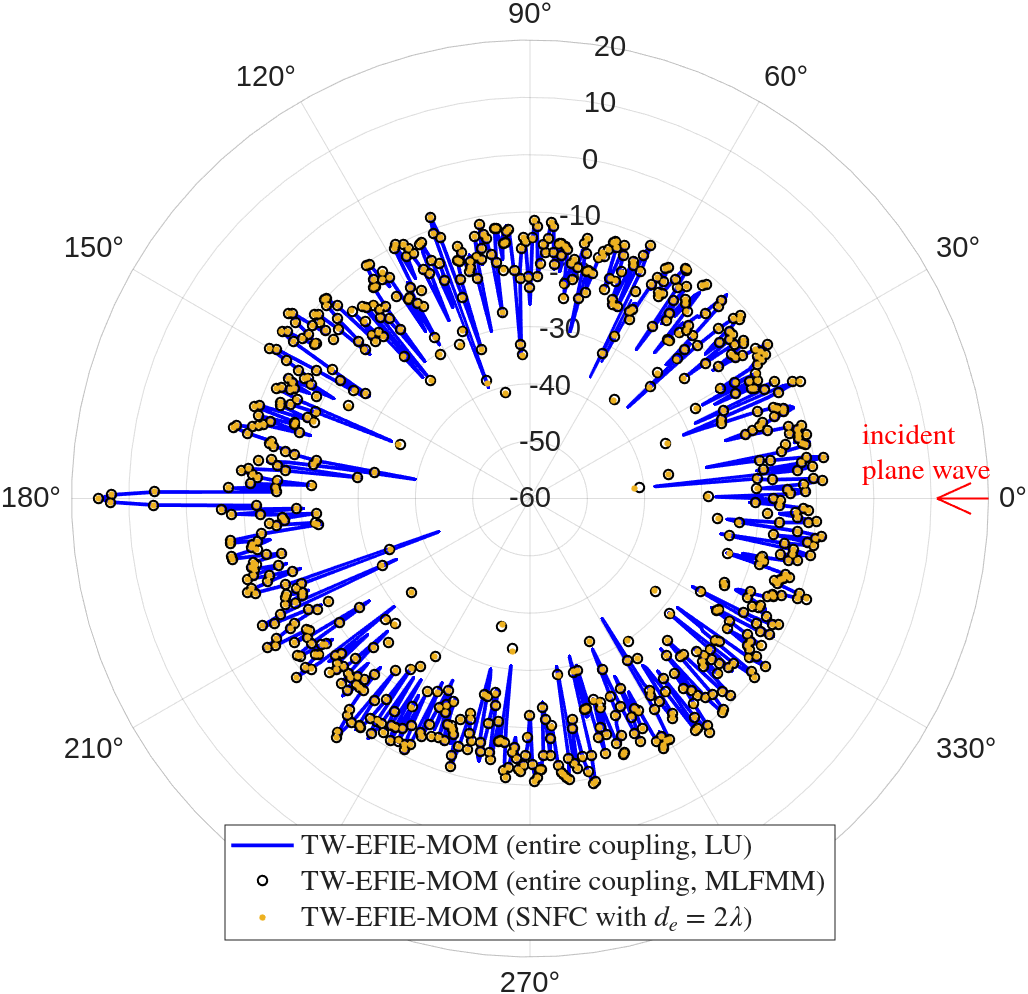}
    \caption*{(b) Bent chaff cloud}
  \end{minipage}
  \hfill
  % Third image
  \begin{minipage}[b]{0.32\textwidth}
    \centering
    \includegraphics[width=\textwidth]{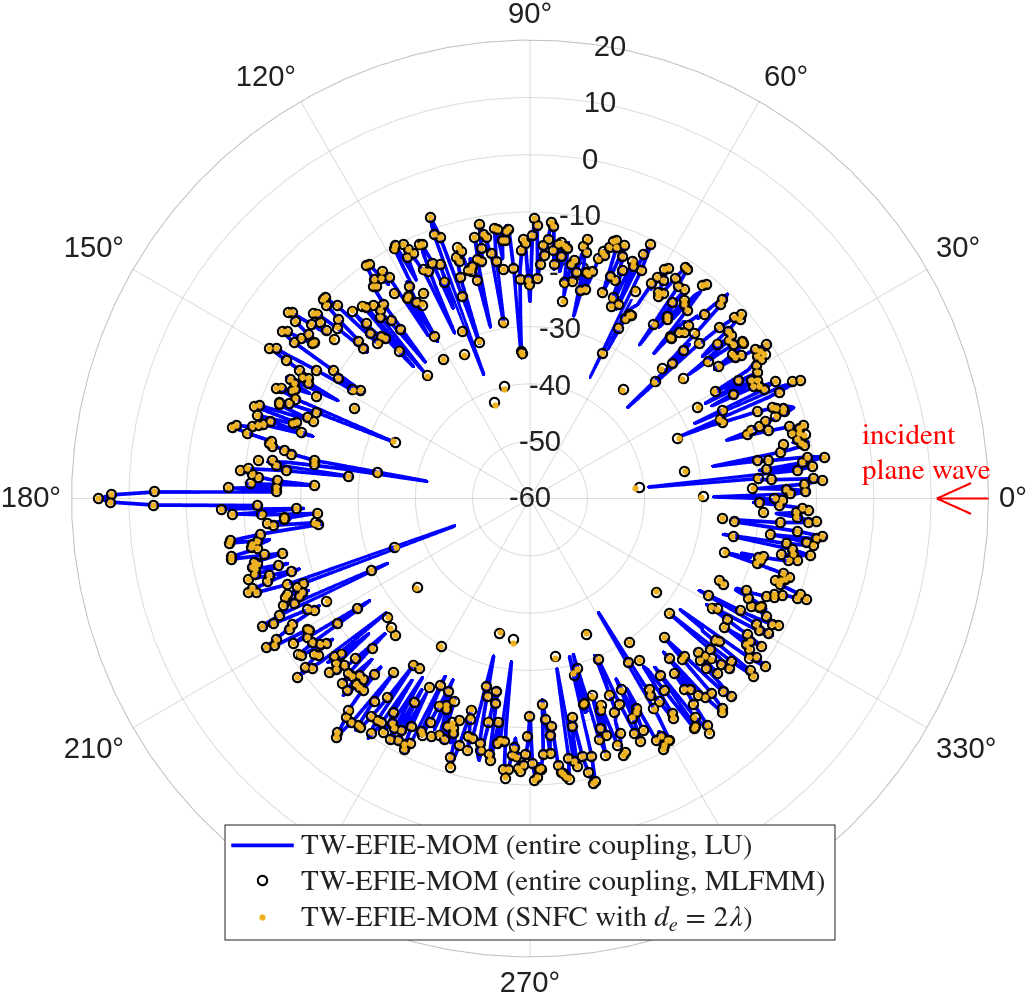}
    \caption*{(c) Twisted bent chaff cloud}
  \end{minipage}
  \caption{Bistatic RCS of three different chaff clouds with $1{,}000$ chaff elements each, computed using LU, MLFMM, and SNFC. The elements are randomly oriented and distributed inside a $0.5$\,m cloud.}
  \label{fig:cloud_rcs}
\end{figure*}

The simulations are performed at an operating frequency of $16.854\,~[\text{GHz}]$, corresponding to a free-space wavelength of $\lambda \approx 17.8\,~[\text{mm}]$.
With $d_{\text{mean}} \approx 4.5\lambda$, the overall cloud exhibits weak but non-negligible coupling among neighboring elements, resulting in a sparse impedance matrix that motivates the SNFC formulation.
The incident field is specified with azimuth angle $\phi_{\text{inc}}=0^{\circ}$, zenith angle $\theta_{\text{inc}}=90^{\circ}$, and V-polarization. 
The bistatic VV-polarized RCS is observed in the plane at $\theta_{\text{obs}}=90^{\circ}$, with azimuthal angle $0^{\circ} \leq \phi_{\text{obs}} \leq 360^{\circ}$ sampled in $0.5^{\circ}$ steps. 
The results obtained using LU, MLFMM, and SNFC methods are shown for each case. 
As illustrated in Fig.~\ref{fig:cloud_rcs}, the SNFC results are greatly consistent with those by LU and MLFMM regardless of the chaff element shape, thereby confirming the validity of the SNFC approximation even for bent and twisted-bent chaff ensembles.

In our implementation, the sparse structure of the SNFC impedance matrix for large-scale problems is organized by adopting the oct-tree data structure commonly used in MLFMM. 
This hierarchical partitioning enables both methods to handle millions of unknowns systematically. 

For both MLFMM and SNFC, we employ a generalized conjugate residual (GCR)~\cite{article:gcr1} iterative solver, which monotonically decreases the residual error with each iteration. 
As a preconditioner, a modified block-diagonal preconditioner (BDP) is applied to the right-preconditioned system, namely,
\begin{equation}
    \overline{\mathbf{Z}}\cdot \overline{\mathbf{P}}\cdot\mathbf{y}= \overline{\mathbf{B}}\cdot \mathbf{y} = \mathbf{v},
\label{eq:preconditioned_system}
\end{equation}
where $\overline{\mathbf{P}}$, $\overline{\mathbf{B}}$, and $\mathbf{y}$ denote the preconditioner matrix, the preconditioned system matrix, and the solution vector of the right-preconditioned system, respectively. 
In~\eqref{eq:preconditioned_system}, the auxiliary variable $\mathbf{y}$ is first solved, and the original solution vector can then be recovered by $\mathbf{j} = \overline{\mathbf{P}}\cdot\mathbf{y}$. 
Conventional BDPs utilize diagonal blocks defined at the leaf level of the oct-tree hierarchy. 
In contrast, our modified BDP constructs block diagonals from one level above the leaves, effectively grouping several daughter blocks into a larger parent block. 
This strategy improves the conditioning of the system while maintaining both sparsity and scalability.
Furthermore, the modified BDP is straightforward to implement, since it directly reuses the near-field interaction matrices that are already constructed in the MLFMM framework. 
In other words, the block structure of the preconditioner naturally aligns with the existing oct-tree partitioning, allowing one to assemble the modified block diagonals without additional near-field evaluations or data structures. 
For the SNFC formulation, the implementation is even simpler because the system matrix itself is inherently sparse and organized in a block-wise manner. 
Therefore, the same concept of grouping near-field blocks can be applied without any hierarchical tree traversal, making the modified BDP particularly attractive for large-scale sparse-matrix systems such as SNFC.
\begin{figure}[t]
  \centering
  \includegraphics[width=0.48\textwidth]{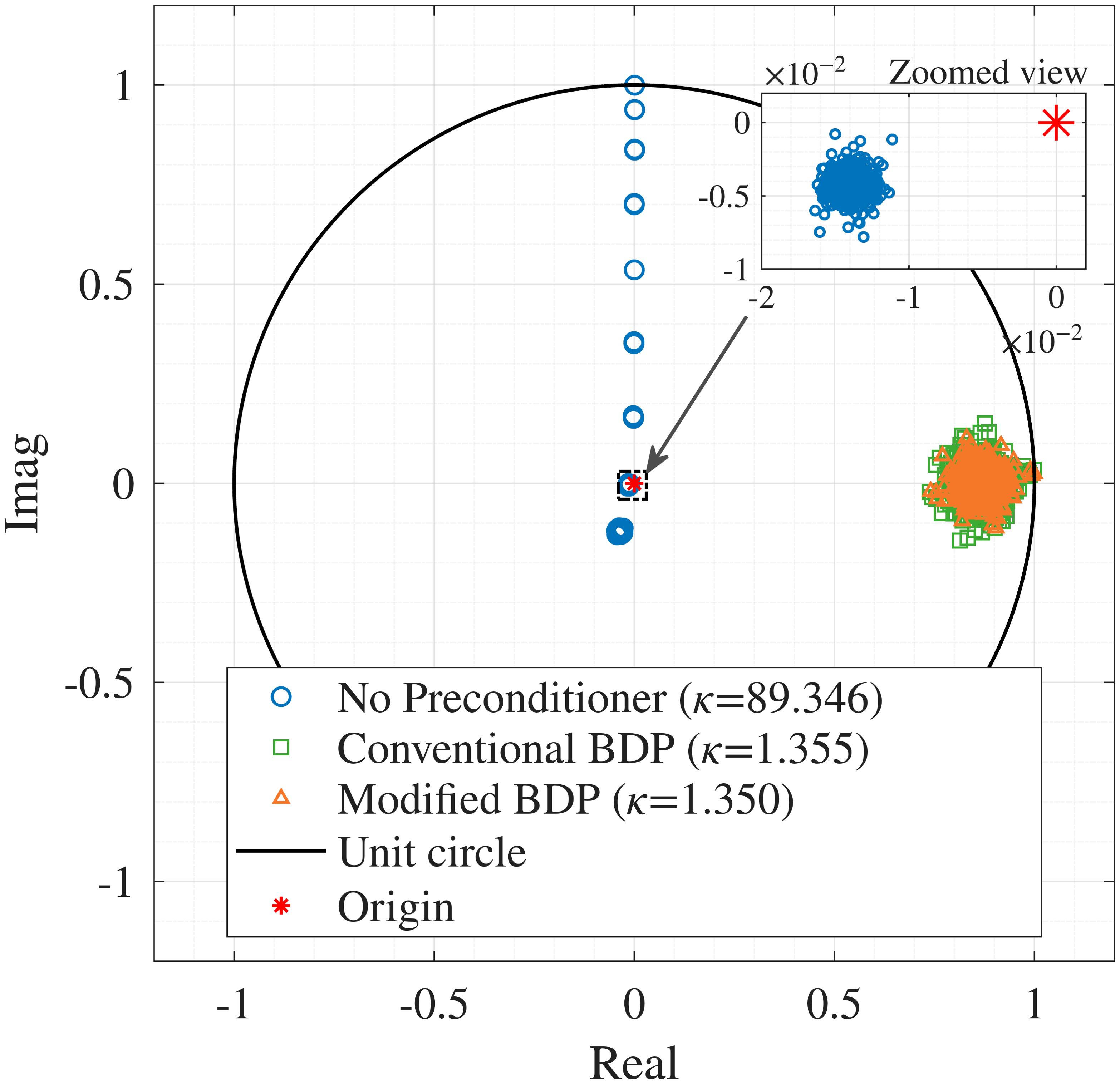}
  \caption{Eigenvalue distribution of SNFC impedance matrix for a cloud of $1{,}000$ chaffs before and after applying modified bock diagonal preconditioner.}
  \label{fig:eigenvalues}
\end{figure}

As shown in Fig.~\ref{fig:eigenvalues}, the normalized eigenvalue distributions are presented to illustrate the spectral effect of the preconditioner. 
Specifically, for each case we plot
\begin{equation}
\overline{\lambda}_i := \frac{\lambda_i}{\text{max}\,|\lambda({\overline{\mathbf{B}}})|},
\end{equation}
and, for the case without a preconditioner, the eigenvalues of the original impedance matrix $\overline{\mathbf{Z}}$ are normalized in the same manner.
The condition number is defined as
\begin{equation}
\kappa = \frac{\text{max} |\lambda ({\overline{\mathbf{B}}})|}{\text{min} |\lambda ({\overline{\mathbf{B}}})|}
\end{equation}
which is significantly reduced by both the conventional BDP and the proposed modified BDP. 
In the un-preconditioned system, the eigenvalues are broadly spread along the imaginary axis, and a subset of them are loosely clustered near the origin, indicating poor spectral conditioning. 
After preconditioning, most of the eigenvalues shift toward the positive real axis and form a concentrated cluster centered approximately at $(0.85, 0)$ with a radius of approximately $0.3$,
demonstrating a tighter spectral spread and improved numerical conditioning of the preconditioned matrix.
\begin{figure}[t]
  \centering
  \includegraphics[width=0.48\textwidth]{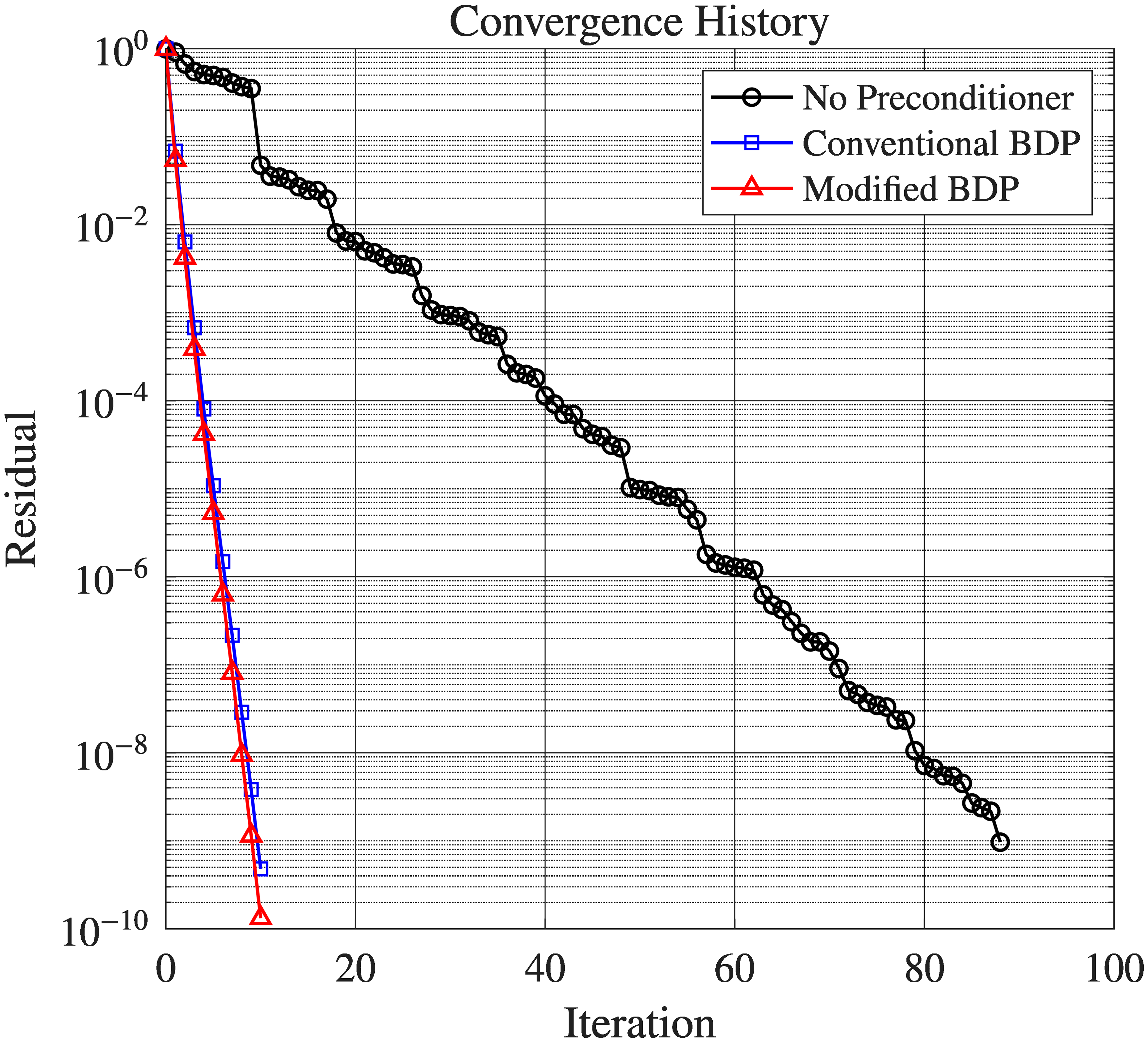}
  \caption{Convergence history for a $0.5$ meter radius cloud of $1{,}000$ chaffs with respect to the preconditioner.}
  \label{fig:convergence_behavior}
\end{figure}
The convergence behavior is reported in Fig.~\ref{fig:convergence_behavior}, where the normalized residual error is defined as
\begin{equation} 
    \epsilon_r=\frac{\left|\mathbf{v}-\overline{\mathbf{B}}\cdot\mathbf{y}_i\right|}{\left|\mathbf{v}\right|},
\end{equation}
and plotted with respect to iteration count for the GCR solver under a $10^{-9}$ stopping criterion, where $\mathbf{y}_i$ is a solution at $i$-th iteration. 
Because the conventional BDP already yields a sufficiently small condition number for this test size, the difference in convergence rate between the conventional and modified BDPs remains modest. 
However, as the number of chaff elements increases, the benefit of the modified BDP becomes evident: by grouping oct-tree blocks one level above the leaves, the preconditioner captures stronger intra-group couplings while preserving sparsity, thereby tightening the spectrum of $\overline{\mathbf{B}}$ and consistently reducing the iteration count for large-scale problems.
%\begin{figure}[t]
%  \centering
%  \includegraphics[width=0.48\textwidth]{Figures/RCS_1k_Sph/convergence_behavior_1mil.png}
%  \caption{Convergence history for a 10 meter radius spherical cloud of $1$ million chaffs with respect to the preconditioner.}
%  \label{fig:convergence_behavior_1mil}
%\end{figure}

\begin{figure}[t]
  \centering
  \includegraphics[width=0.35\textwidth]{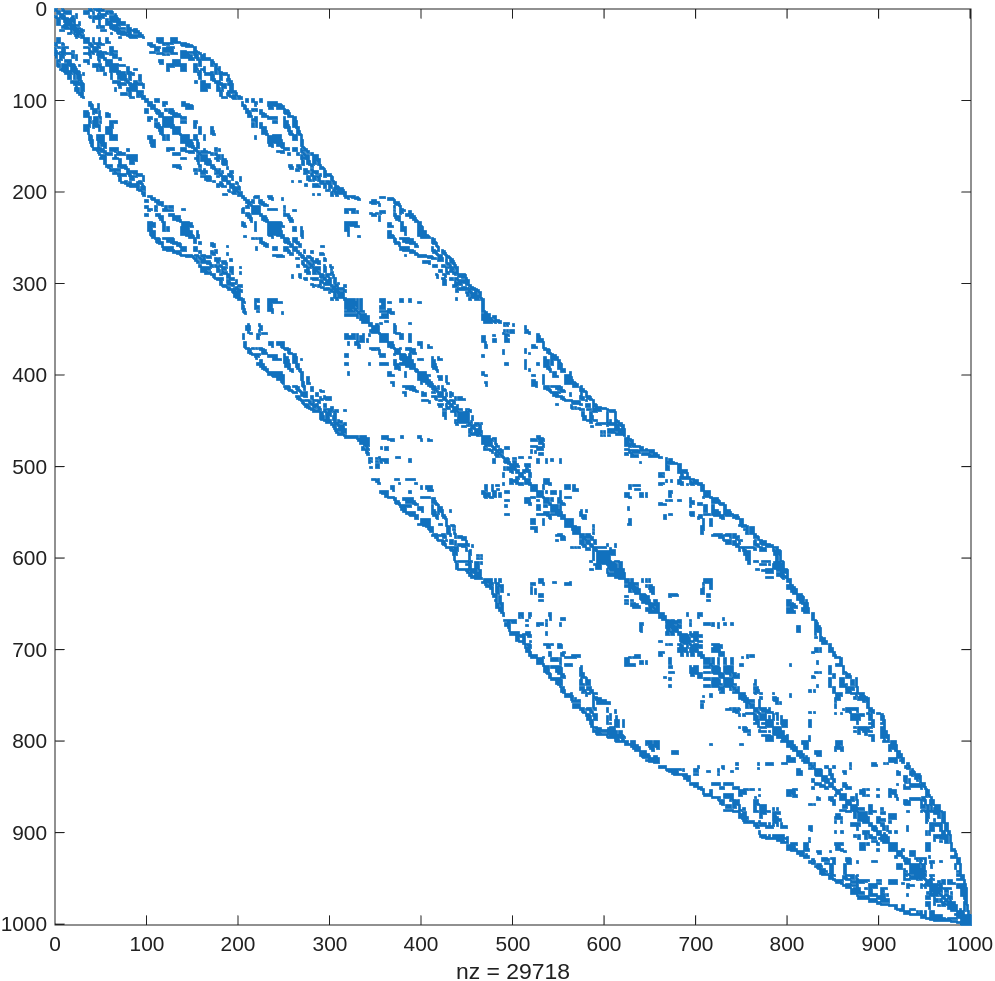}
  \caption{Sparsity pattern of the SNFC impedance matrix for a cloud of $1{,}000$ chaffs. Each nonzero entry corresponds to a $9 \times 9$ block. The matrix is reordered using the Reverse Cuthill--McKee algorithm~\cite{RCM}.}
  \label{fig:sparsity}
\end{figure}

In addition, the sparsity pattern of the SNFC impedance matrix is analyzed for the same case of 1,000 chaff elements. 
Although the full impedance matrix has a dimension of $9{,}000 \times 9{,}000$, the large inter-chaff separation ensures that the overall sparsity pattern closely resembles that of a single chaff element. 
Each visible entry in Fig.~\ref{fig:sparsity} corresponds to a $9 \times 9$ subblock. 
To further elucidate the matrix structure, the Reverse Cuthill--McKee algorithm~\cite{RCM} is applied to reorder the matrix, revealing the banded sparsity structure characteristic of the SNFC formulation. 
The average number of nonzero elements per row is $29.7$, indicating that each chaff interacts with approximately $30$ neighboring chaffs in this case.
It is clearly observed that the use of SNFC acceleration results in a significantly sparser coupling compared to a conventional MoM solver, which constitutes the core computational advantage of THEM-S and enables efficient RCS estimation for million-scale chaff cloud configurations.

\section{Real-Time monostatic RCS Estimation for a Freely Falling Chaff Cloud Composed of One Million Chaffs}\label{sec7}
In this section, the developed aerodynamic--electromagnetic coupled module is employed to simulate the time evolution of the monostatic RCS of a chaff cloud consisting of one million chaff elements during free fall. 
As demonstrated in the previous examples, the spatial distribution and orientation of a chaff cloud vary significantly during descent depending on the chaff geometry. 
To comparatively analyze these effects, one-million-chaff clouds composed of S-, B-, and TB-type chaffs are simulated using the present 6-DOF formulation to capture the aerodynamic behavior. 
The monostatic RCS is then computed over time at $\theta_{\mathrm{inc}} = 90^{\circ}$ and $\phi_{\mathrm{inc}} = 0^{\circ}$ for comparison.

To model the free-fall behavior of a chaff cloud immediately after deployment, the cloud is assumed to consist of one million chaff elements uniformly distributed within a spherical region of radius $10~[\mathrm{m}]$ centered at $(0,\,0,\,100)~[\mathrm{m}]$. 
The initial orientations of the chaffs are treated as random variables following a normal distribution with a mean elevation angle of $0^{\circ}$ and a standard deviation of $30^{\circ}$. 
The chaff diameters are also modeled as random variables with a mean of $27.01~[\mu\mathrm{m}]$ and a standard deviation of $2.98~[\mu\mathrm{m}]$, while the corresponding masses follow a normal distribution with a mean of $27.17~[\mu\mathrm{g}]$ and a standard deviation of $6.90~[\mu\mathrm{g}]$. 
All chaffs have an identical length of $L = 1.78~[\mathrm{cm}]$. 

The chaff dynamics are simulated for a total duration of $1~[\mathrm{min}]$, and the monostatic VV- and HH-polarized RCS values are computed every $0.1~[\text{sec}]$. 
Zero initial translational and angular velocities are assumed. 

In modeling the TB-type chaff cloud, each chaff element is assigned a bending radius $R_b$ and a twisting amplitude $A_t$ that incorporate random geometric variations. 
The bending radius $R_b$ follows a uniform distribution with a mean of $10.7~[\mathrm{cm}]$ and a range of $[1.78,\,19.58]~[\mathrm{cm}]$, while the twisting amplitude $A_t$ has a mean of zero and a standard deviation of approximately $0.1\times1.78~[\mathrm{cm}] = 0.178~[\mathrm{cm}]$. 
These variations introduce small geometric perturbations to the twisted--bent chaff shape, effectively accounting for irregularities arising from fabrication and deployment processes. 
For comparison, the S-type chaff is modeled with a very large $R_b$ (on the order of $10^5$) and $A_t = 0$, whereas the B-type chaff is simulated with $A_t = 0$. 

To reduce RCS fluctuations and capture the overall scattering pattern, 6 independent aerodynamic--electromagnetic coupled simulations for each chaff geometry are performed with different random initial conditions. 
The VV- and HH-polarized monostatic RCSs are then obtained by averaging the RCS values across all realizations at each time step.

\begin{figure*}[t]
    \centering
    \begin{subfigure}{0.32\linewidth}
        \centering
        \includegraphics[width=\linewidth]{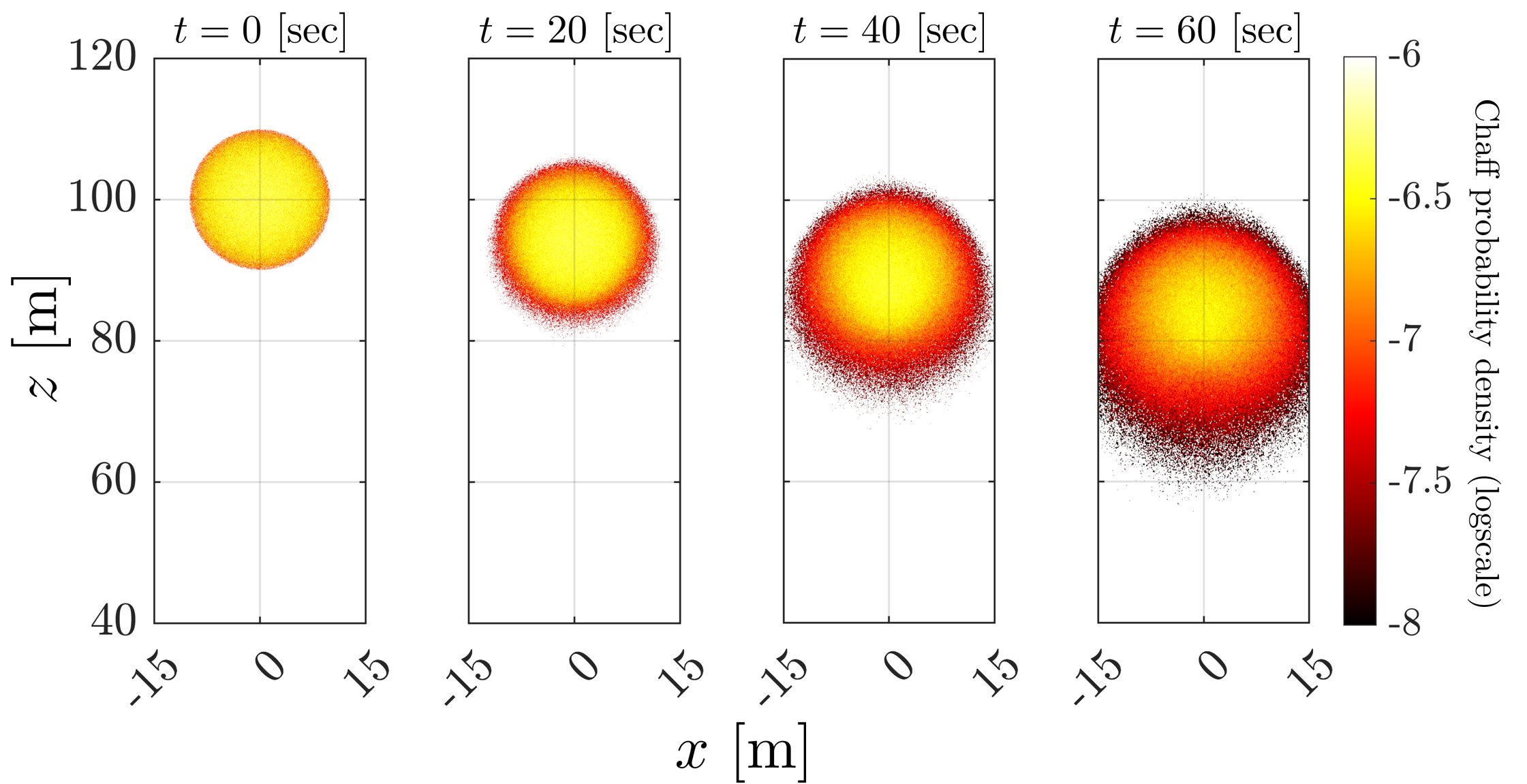}
        \caption{S-type chaff cloud.}
        \label{fig:S_pdf}
    \end{subfigure}
    \hfill
    \begin{subfigure}{0.32\linewidth}
        \centering
        \includegraphics[width=\linewidth]{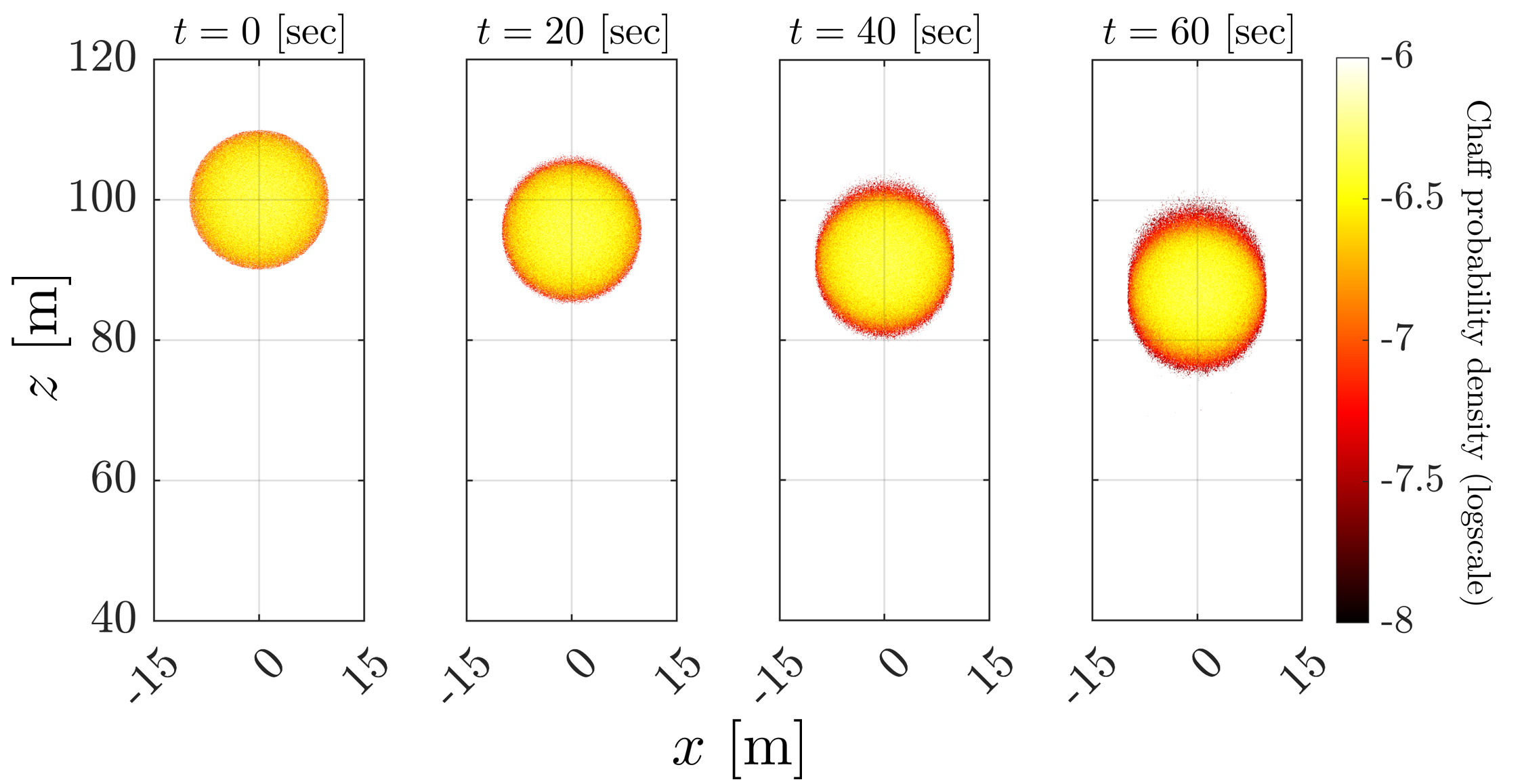}
        \caption{B-type chaff cloud.}
        \label{fig:B_pdf}
    \end{subfigure}
    \hfill
    \begin{subfigure}{0.32\linewidth}
        \centering
        \includegraphics[width=\linewidth]{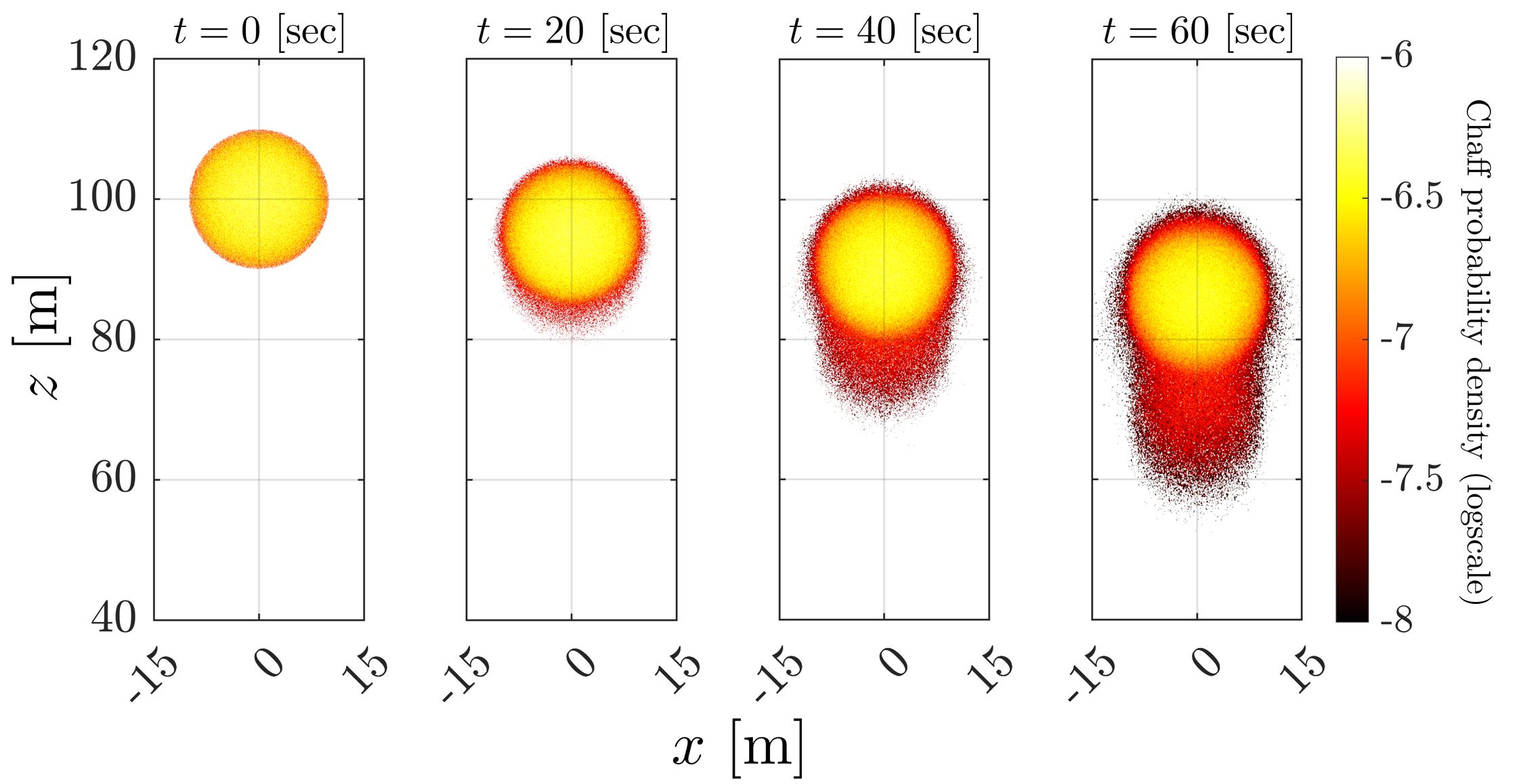}
        \caption{TB-type chaff cloud.}
        \label{fig:TB_pdf}
    \end{subfigure}
    \caption{Time-evolving probability density (log scale) of the chaff cloud in the $x$--$z$ plane for (a) S-type, (b) B-type, and (c) TB-type cases.}
    \label{fig:chaff_1e6_pdf}
\end{figure*}

Fig.~\ref{fig:chaff_1e6_pdf} illustrates the time-evolving probability density (log scale) of the chaff cloud in the $x$–$z$ plane for (a) S-type, (b) B-type, and (c) TB-type cases. 
In the S-type case, the chaff cloud is observed to continuously expand over time. 
This occurs because the S-type chaff, owing to its geometric symmetry, experiences no external aerodynamic moment and thus preserves its initial orientation throughout the descent. 
Depending on the initial attitude, each chaff element experiences different aerodynamic forces along the axial and normal directions. 
When oriented vertically, the chaff falls rapidly without side slip, whereas as it gradually transitions toward a horizontal orientation, the terminal velocity decreases and side-slip motion emerges. 
As a result, even in the absence of external wind or turbulence, the chaff cloud expands significantly over time. 
Although this behavior appears physically reasonable, it represents an idealized condition; in actual experiments, such continuous expansion of the chaff cloud has not been observed.

On the other hand, in the B-type case, the overall size of the chaff cloud remains nearly constant over time. 
This occurs because the two-dimensionally bent B-type chaffs rapidly settle into a flattened orientation under the influence of the aerodynamic damping moment during free fall. 
Since the terminal velocity in this flattened state is smaller than that in the vertical orientation, the chaffs descend more slowly.

In the TB-type case, the chaff cloud maintains a cylindrical radial extent similar to that of the B-type case; however, its vertical distribution gradually broadens over time. 
As demonstrated in the previous example, the three-dimensionally twisted--bent chaffs exhibit not only flattened orientations but also helical motions. 
Because the onset of helical motion varies depending on the degree of bending and the initial attitude, the chaff cloud exhibits altitude-dependent broadening of its spatial distribution as time progresses.

\begin{figure*}[t]
    \centering
    \begin{subfigure}{0.32\linewidth}
        \centering
        \includegraphics[width=\linewidth]{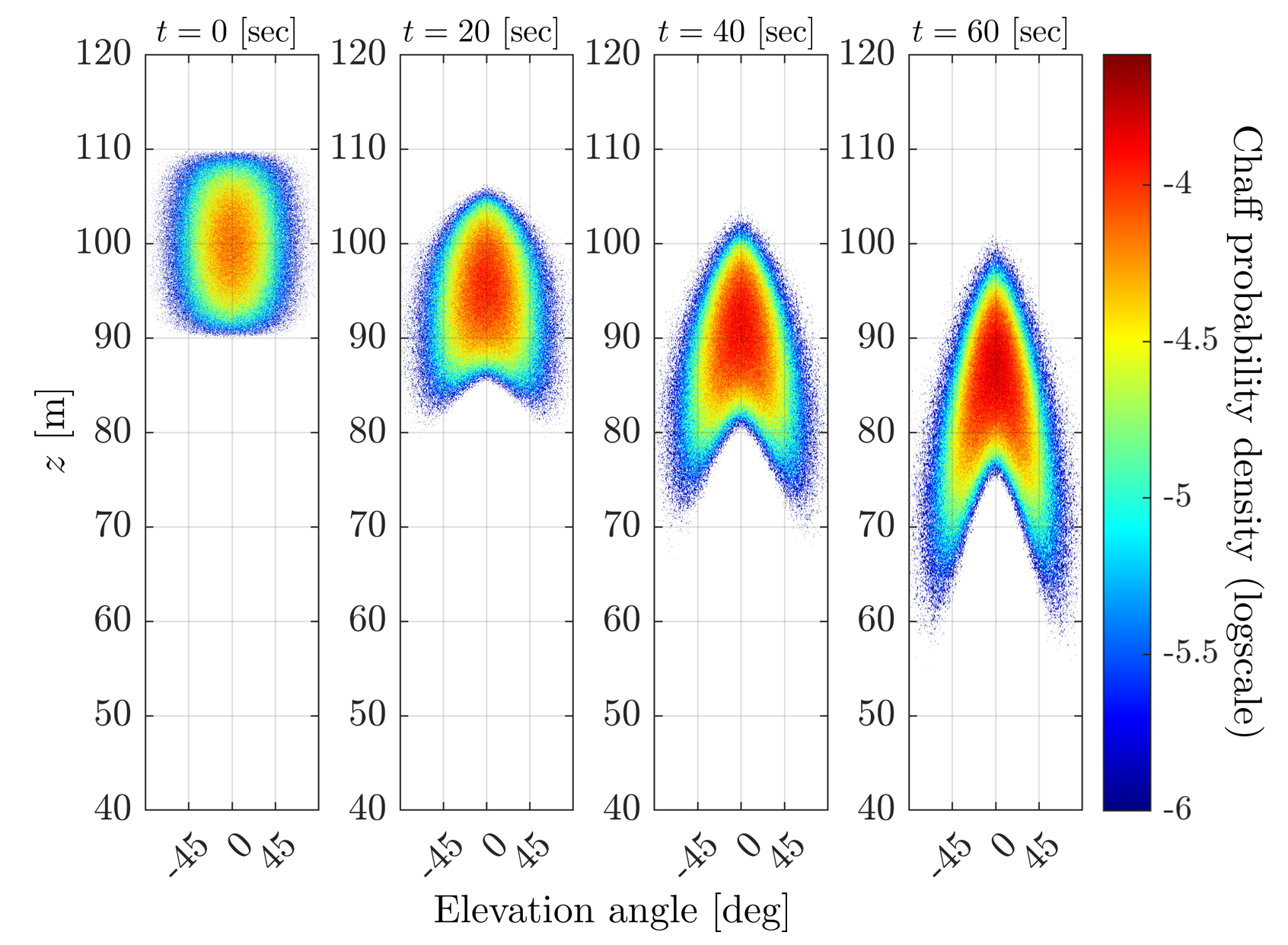}
        \caption{S-type chaff cloud.}
        \label{fig:S_ori_pdf}
    \end{subfigure}
    \hfill
    \begin{subfigure}{0.32\linewidth}
        \centering
        \includegraphics[width=\linewidth]{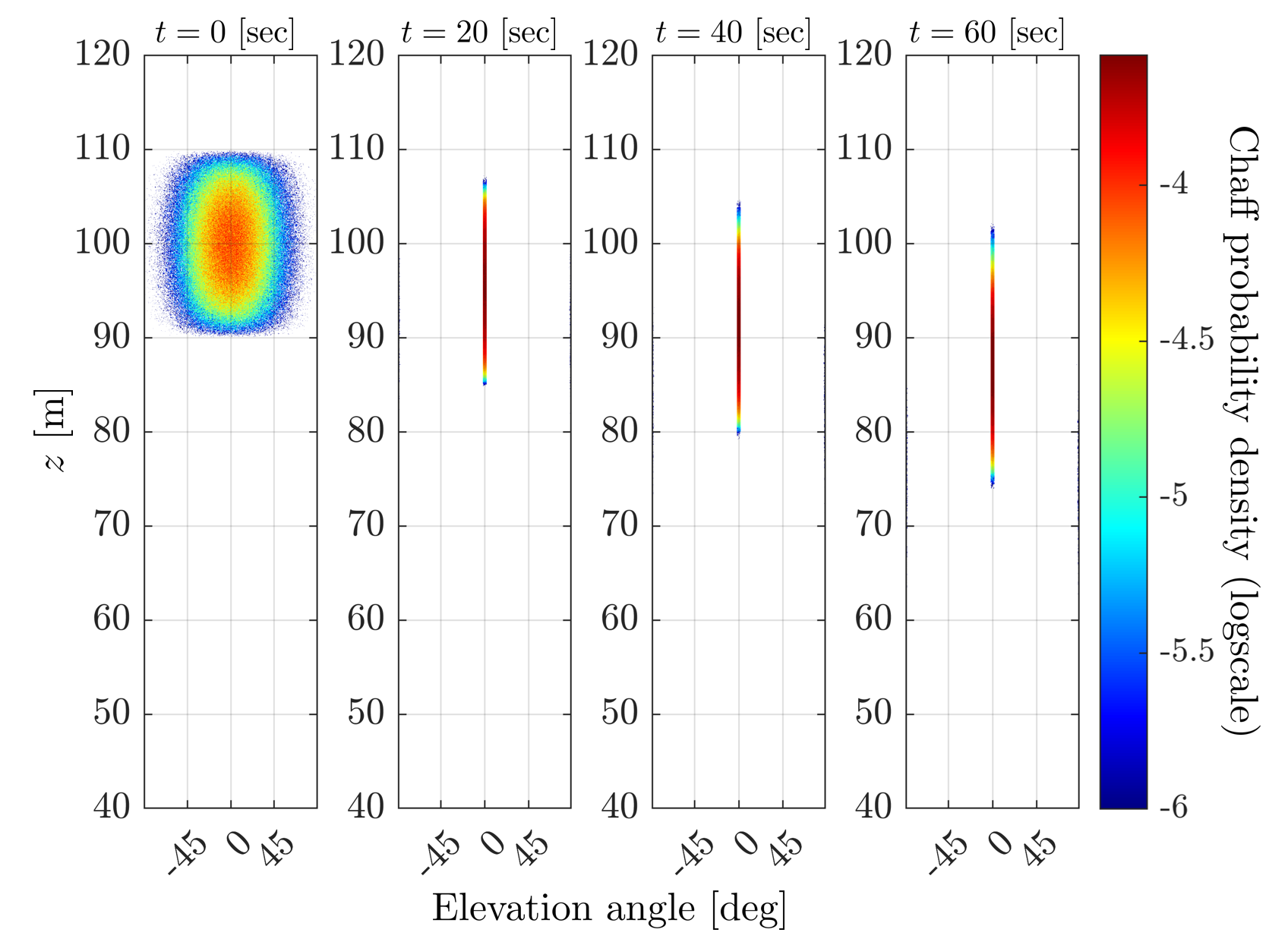}
        \caption{B-type chaff cloud.}
        \label{fig:B_ori_pdf}
    \end{subfigure}
    \hfill
    \begin{subfigure}{0.32\linewidth}
        \centering
        \includegraphics[width=\linewidth]{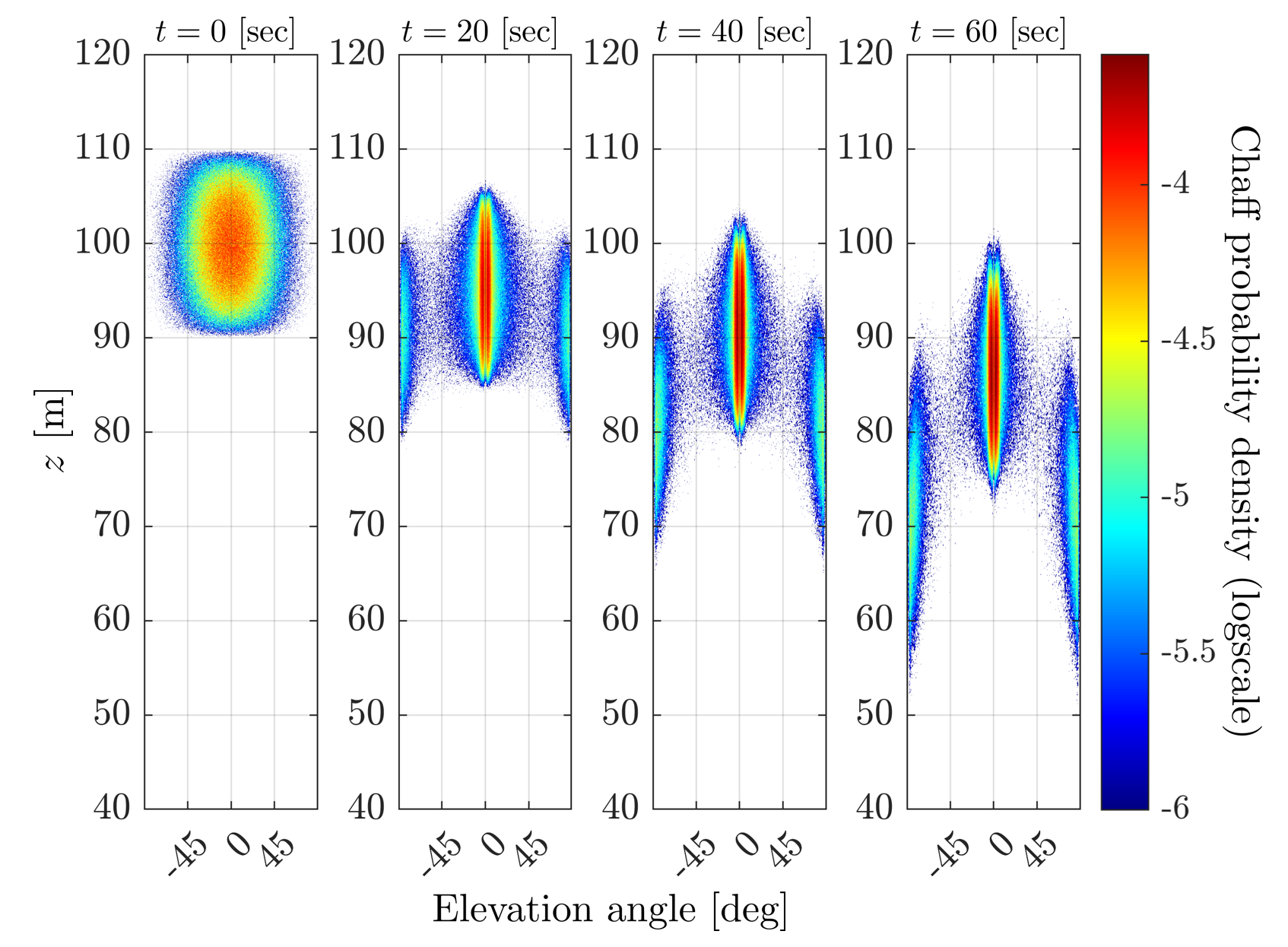}
        \caption{TB-type chaff cloud.}
        \label{fig:TB_ori_pdf}
    \end{subfigure}
    \caption{Probability density (log scale) of chaff orientation as a function of altitude at selected time points for (a) S-type, (b) B-type, and (c) TB-type cases.}
    \label{fig:aero_1e6_orientation}
\end{figure*}
\begin{figure*}[t]
  \centering
  \begin{subfigure}{0.32\linewidth}
      \centering
      \includegraphics[width=\linewidth]{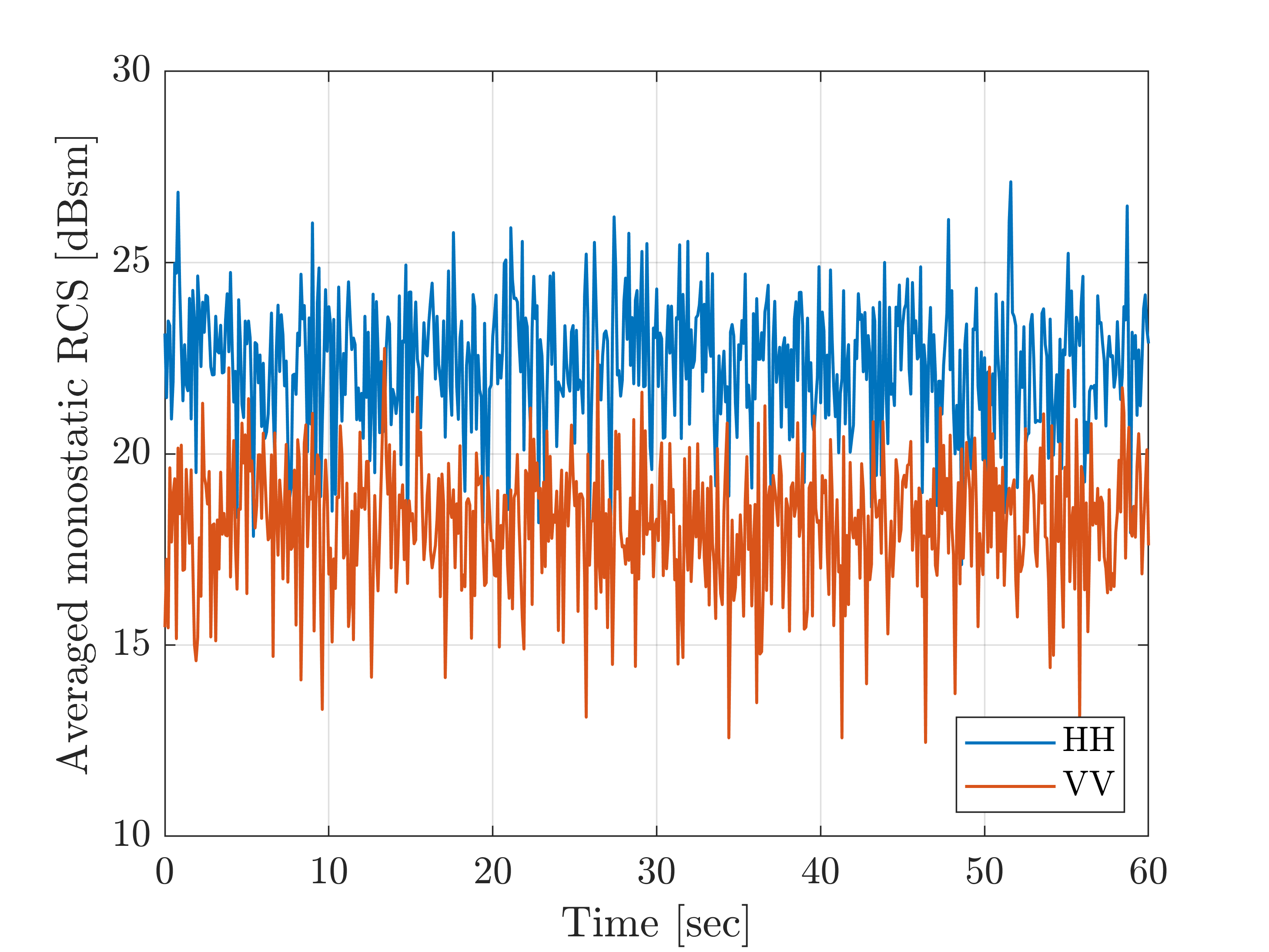}
      \caption{S-type chaff cloud.}
      \label{fig:S_RCS}
  \end{subfigure}
  \hfill
  \begin{subfigure}{0.32\linewidth}
      \centering
      \includegraphics[width=\linewidth]{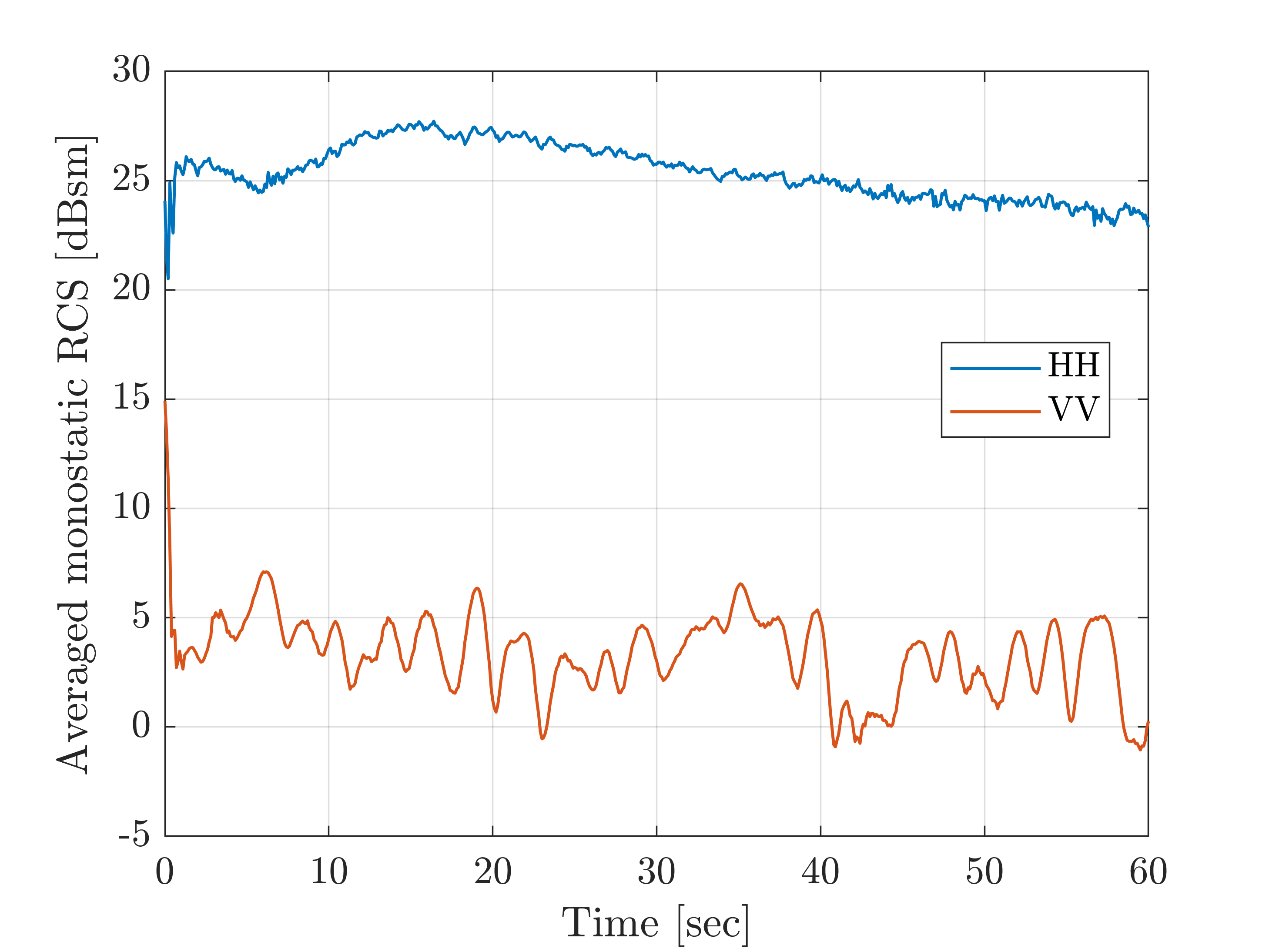}
      \caption{B-type chaff cloud.}
      \label{fig:B_RCS}
  \end{subfigure}
  \hfill
  \begin{subfigure}{0.32\linewidth}
      \centering
      \includegraphics[width=\linewidth]{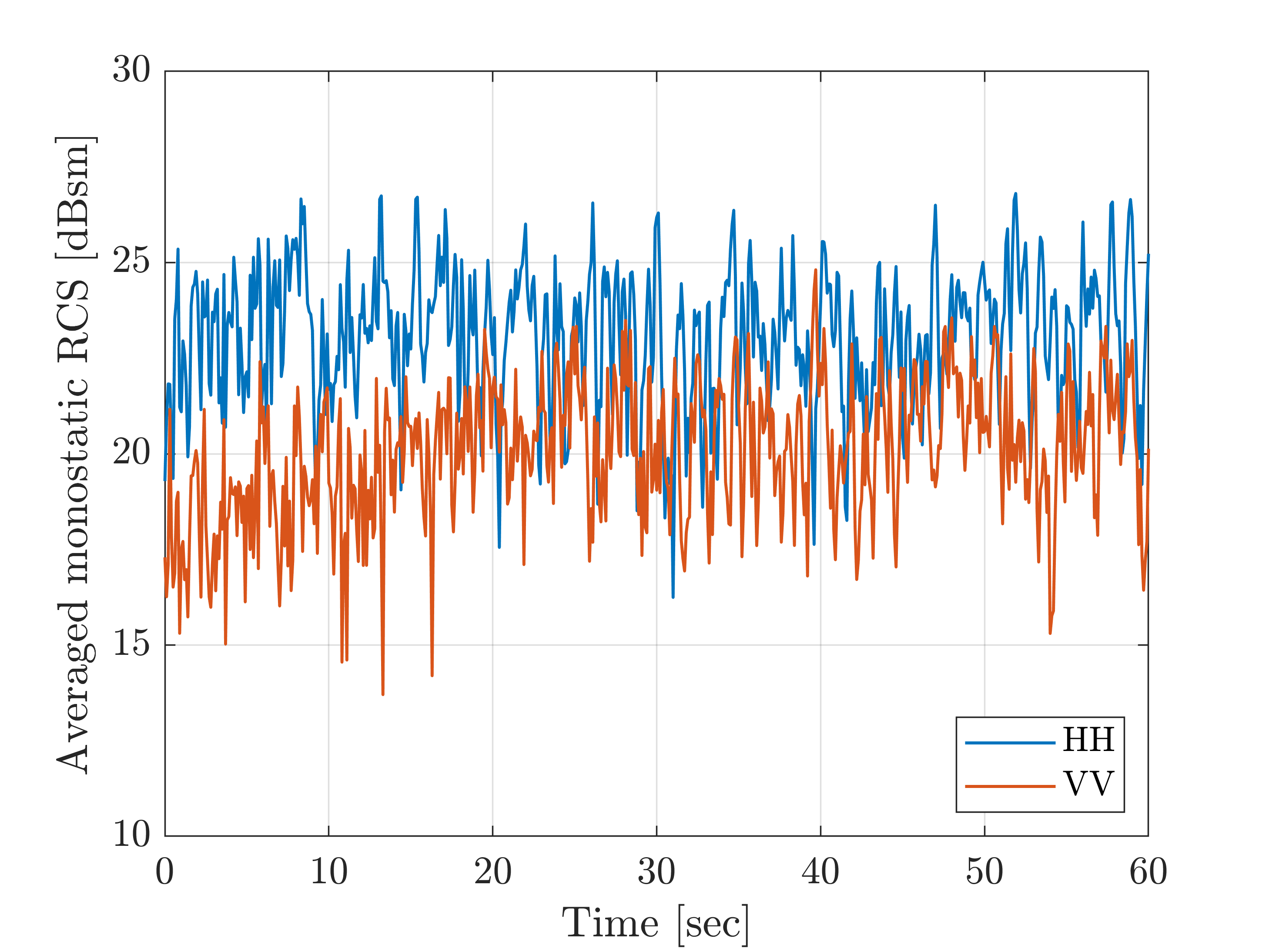}
      \caption{TB-type chaff cloud.}
      \label{fig:TB_RCS}
  \end{subfigure}
  \caption{Comparison of the monostatic RCS of S-, B-, and TB-type chaff clouds.}
  \label{fig:1e6_RCS}
\end{figure*}

To examine the time evolution of the chaff cloud orientation, the elevation probability density function (PDF) of chaff elements at different altitudes is plotted in Figs.~\ref{fig:aero_1e6_orientation} for the three chaff types.
The S-type chaff maintains its initial orientation distribution; however, since vertically oriented chaff falls faster than horizontally oriented ones, the overall distribution gradually forms a spherical shell shape over time, with orientations ranging from horizontal near the equator to vertical near the poles.
For the B-type chaff, most elements rapidly converge to a horizontal orientation within a few seconds.
In contrast, the TB-type chaff exhibits a bimodal distribution over time, showing both horizontally aligned and vertically helical falling modes.

For the three types of chaff clouds, the VV- and HH-polarized monostatic RCS values were computed at intervals of 0.1~[sec], as shown in Fig.~\ref{fig:1e6_RCS}. 
To mitigate the influence of random initial conditions on the RCS characteristics, seven independent free-fall simulations were performed for each chaff geometry.  
The monostatic RCS was evaluated as a function of time for each realization, and the results were subsequently averaged over all realizations to obtain the time-dependent mean RCS.

In the B-type case, most chaff elements rapidly converge to a flattened orientation, resulting in a transient increase in $\sigma_{HH}$ and a corresponding decrease in $\sigma_{VV}$ within the first 2~[sec].  
The local minimum observed around 6~[sec] and the gradual monotonic decrease of $\sigma_{HH}$ after 20~[sec] are attributed to insufficient statistical averaging, reflecting the residual influence of the random initial conditions, particularly the azimuthal angle.  
Overall, $\sigma_{VV}$ remains approximately 23~[dB] lower than $\sigma_{HH}$, which is inconsistent with experimental observations.  
Previous studies have reported that the difference between HH- and VV-polarized monostatic RCS is typically less than 10~[dB]~\cite{arnott2004radar,marcus2015bistatic,seo2012dynamic,brunk1975chaff}.

In the S-type case, the RCS difference remains around 5~[dB] throughout the simulation period, which appears reasonable but merely reflects the initial orientations of the chaff elements.  
Hence, the S-type model does not reproduce realistic experimental behavior.  
In contrast, the TB-type chaff cloud exhibits a similar difference of approximately 5~[dB], which is physically plausible.  
More importantly, this difference gradually decreases up to about 20~[sec], as some chaff elements begin to exhibit helical motion at later stages.  
Consequently, $\sigma_{VV}$ slowly increases with time and saturates after approximately 20~[sec], whereas $\sigma_{HH}$ remains nearly constant or slightly decreases.  
Such an increase in the VV-polarized monostatic RCS has also been reported in previous studies~\cite{arnott2004radar,seo2012dynamic,marcus2004dynamics}.  
Therefore, the present coupled aerodynamic--electromagnetic module, incorporating an arbitrary three-dimensional chaff geometry model, successfully reproduces the realistic free-fall behavior of chaff clouds.

\section{Conclusion and Future Works}\label{sec8}
In this study, we presented a coupled electromagnetic–aerodynamic modeling framework capable of predicting the radar cross section (RCS) of large-scale chaff clouds from first principles.  
By extending the conventional six-degree-of-freedom (6-DoF) formulation to incorporate arbitrarily curved three-dimensional chaff geometries, the proposed method accurately captures the aerodynamic moments responsible for both flattened and helical motions.  
The coupling between the aerodynamic solver and the fast method-of-moments solver, THEM-S, enables real-time RCS prediction for chaff clouds consisting of up to one million elements.  
The simulation results confirm that the RCS evolution of a chaff cloud is strongly governed by its aerodynamic behavior.  
For instance, in the B-type case, most chaffs converge rapidly to flattened orientations, resulting in a large polarization-dependent RCS difference exceeding 20~dB, which is inconsistent with experimental observations.  
In contrast, the TB-type case---featuring three-dimensionally twisted--bent geometries---reproduces more realistic free-fall dynamics, maintaining a small VV-HH difference of only a few decibels.  
Moreover, as the simulation progresses, the VV-polarized monostatic RCS gradually increases due to the delayed onset of helical motion among individual chaff elements, while the HH component remains nearly constant. 
These results demonstrate that the proposed framework can not only reproduce experimentally observed RCS behavior but also provide physically grounded insight into the dynamic scattering mechanisms of large-scale chaff clouds.  
The present approach therefore establishes a foundation for future extensions toward data-driven statistical modeling and validation using measurement campaigns.

In practice, it is extremely challenging to measure the time-varying RCS of chaff clouds, 
since the cloud rapidly disperses and evolves in both spatial and angular distributions after deployment. 
Moreover, the short radar dwell time and the loss of phase coherence between successive radar pulses 
make it practically impossible to obtain directly time-resolved RCS measurements. 
Therefore, numerical simulation provides an essential means to analyze and predict the temporal evolution of chaff RCS under controlled aerodynamic conditions.
The time evolution of chaff RCS estimated using the proposed aerodynamic--electromagnetic coupled module can provide valuable insights for radar signal processing, particularly in reconstructing Range–Doppler (RD), Range–Angle (RA), and Doppler–Angle (DA) maps. 
In future work, we plan to extend the present framework to compute the complex far-field responses 
of chaff clouds at each time frame and frequency, and to develop a radar signal processing methodology 
that enables realistic numerical synthesis of RD, RA, and DA maps based on these time-dependent electromagnetic responses.

%%%%%%%%%%%%%%%%%%%%%%%%%%%%%%%%%%%%%%%%%%%%%%%%%%%%%%%%%%%%%%%%%%%%%%%%%%%%%%%%%%%%%%

\ifCLASSOPTIONcaptionsoff
\newpage
\fi

%\section{Appendix}

\bibliographystyle{IEEEtran}
\bibliography{IEEEabrv,mybib}

\end{document}